\documentclass[aps,showpacs,preprint]{revtex4}
\usepackage{graphicx,color}% Include figure files
\usepackage{dcolumn}% Align table columns on decimal point
\usepackage{bm}% bold math
\usepackage{amsmath,amsthm,amssymb}

\begin{document}

\title{
Theory of spin relaxation torque in metallic ferromagnets 
}

\author{Noriyuki Nakabayashi, Akihito Takeuchi, Kazuhiro Hosono,  Katsuhisa Taguchi, Gen Tatara}
\affiliation{
Department of Physics, Tokyo Metropolitan University,
Hachioji, Tokyo 192-0397, Japan
}
\date{\today}

\begin{abstract}
Spin transport driven by an external electric field in uniform metallic ferromagnets with the spin-orbit interaction arising from random impurities is studied microscopically.
Spin relaxation torque  ${\cal T}$ is shown to be written by spatial derivatives of 
the electric field, but with anisotropy arising from the magnetization.
The field-driven contribution of the spin current is also anisotropic.
The diffusive spin current is shown to be written as a gradient of the spin chemical potential, 
and the linear-response expression for the spin chemical potential is derived.
It is discussed that the $\beta$ term  in the spin transfer torque can also be anisotropic.
%The spin chemical potential, which drives the diffusive spin curret, is obtained as a function of the applied electric field.
\end{abstract}

\maketitle

%%%%%%%%%%%%%%%%%%%%%%%%%%%%%
%\renewcommand{\half}{\frac{1}{2}} % defs for elsart class
%\newcommand{\half}{\frac{1}{2}} % defs for other class
%%%%%%%%%%%%%%
\def\average#1{\left\langle {#1} \right\rangle}
\def\averagefr#1{\left\langle {#1} \right\rangle_0}
\def\avbar#1{\overline{#1}}
\def\impaverage#1{\left\langle {#1} \right\rangle_{\rm i}}
\def\bra#1{\lt\langle{#1} \rt|}
\def\ket#1{\lt|{#1} \rt\rangle}
\def\braket#1#2{\lt.\lt\langle{#1} \rt|{#2}\rt\rangle}
\def\ddelo#1{\frac{d^2}{d #1^2}}
\def\ddel#1#2{\frac{d^2 #1}{d #2 ^2}}
\def\ddelpo#1{\frac{\partial^2}{\partial #1^2}}
\def\delo#1{\frac{d}{d #1}}
\def\delpo#1{\frac{\partial}{\partial #1}}
\def\deldo#1{\frac{\delta}{\delta #1}}
\def\del#1#2{\frac{d #1}{d #2}}
\def\delp#1#2{\frac{\partial #1}{\partial #2}}
\def\deld#1#2{\frac{\delta #1}{\delta #2}}
\def\vvec#1{\stackrel{{\leftrightarrow}}{#1}} %\def\vvec#1{\overleftrightarrow{#1}}
\def\vectw#1#2{\left(\begin{array}{c} #1 \\ #2 \end{array}\right)}
\def\vecth#1#2#3{\left(\begin{array}{c} #1 \\ #2 \\ #3 \end{array}\right)}
\def\mattw#1#2#3#4{\left(\begin{array}{cc} #1 & #2 \\ #3 & #4 \end{array}\right)}
\def\Eqref#1{Eq. (\ref{#1})}
\def\Eqsref#1#2{Eqs. (\ref{#1})(\ref{#2})}
\def\Eqrefj#1{(\ref{#1})��}
\def\Eqsrefj#1#2{�� (\ref{#1})(\ref{#2})}
\def\dispeq#1{$\displaystyle{#1}$}
\def\ret{{\rm r}}
\def\adv{{\rm a}}
\def\L{{\rm L}}
\def\R{{\rm R}}
\def\green#1#2{g_{#1}^{#2}}
\def\jspin#1#2{j_{{\rm s}#1}^{#2}}
\def\jvspin#1#2{{\jv}_{{\rm s}#1}^{#2}}
\def\ls#1{\ell_{{\rm s}#1}}
%%%%%%%%%%%%
\def\listitem#1{\begin{itemize}\item #1 \end{itemize}}
\newcommand{\lt}{\left}
\newcommand{\rt}{\right}
\newcommand{\nablarl}{\stackrel{\leftrightarrow}{\nabla}}
\newcommand{\nnr}{\nonumber\\}
%%%%%%%%%%%%%%%%%%%%%%%%%%%%%%%%%%%%%%%%%%%%
\newcommand{\adag}{{a^{\dagger}}}
\newcommand{\alphaz}{\alpha_0}
\newcommand{\alphasf}{\alpha_\spinflip}
\newcommand{\Area}{A}
\newcommand{\aT}{\overline{\rm T}}
\newcommand{\av}{{\bm a}}
\newcommand{\Av}{{\bm A}}
\newcommand{\Avem}{{\bm A}_{\rm em}}
\newcommand{\Aem}{{A}_{\rm em}}
\newcommand{\Ams}{{\rm A/m}^2}
\newcommand{\Aph}{A^{\phi}}
\newcommand{\Ath}{A^{\theta}}
\newcommand{\Aphv}{\Av^{\phi}}
\newcommand{\Athv}{\Av^{\theta}}
\newcommand{\Az}{{A^{z}}}
\newcommand{\bv}{{\bm b}}
\newcommand{\Bv}{{\bm B}}
\newcommand{\Bz}{{B_z}}
\newcommand{\Bc}{B_{\rm c}}
\newcommand{\Bvs}{{\bm B}_{S}}
\newcommand{\Bveff}{{\bm B}_{\rm eff}}
\newcommand{\Bve}{{\bm B}_{\rm e}}
\newcommand{\betasf}{{\beta_\spinflip}}
\newcommand{\betana}{{\beta_{\rm na}}}
\newcommand{\betaw}{{\beta_{\rm w}}}
\newcommand{\cbar}{\avbar{c}}
\newcommand{\cdag}{{c^{\dagger}}}
\newcommand{\chiz}{\chi^{(0)}}
\newcommand{\chio}{\chi^{(1)}}
\newcommand{\chitilo}{\tilde{\chi}^{(1)}}
\newcommand{\chitilz}{\tilde{\chi}^{(0)}}
\newcommand{\chiuni}{\chi_{0}}
\newcommand{\cH}{c_{H}}
\newcommand{\cHdag}{c_{H}^{\dagger}}
\newcommand{\ckv}{c_{\kv}}
\newcommand{\ckvs}{c_{\kv\sigma}}
\newcommand{\ccv}{{\bm c}}
\newcommand{\Cbeta}{C_\beta}
\newcommand{\Cr}{C_\rightarrow}
\newcommand{\Cl}{C_\leftarrow}
\newcommand{\Ci}{C_{\rm i}}
\newcommand{\Ct}{C_t}
\newcommand{\cv}{{\bm c}}
\newcommand{\Cv}{{\bm C}}
\newcommand{\ctil}{\tilde{c}}
\newcommand{\dbar}{\avbar{d}}
\newcommand{\deltaS}{\delta S}
\newcommand{\Deltasd}{\Delta_{sd}}
\newcommand{\dels}{{s_0}}
\newcommand{\ddagg}{d^{\dagger}}
\newcommand{\dtil}{\tilde{d}}
\newcommand{\dv}{{\bm d}}
\newcommand{\dw}{{\rm w}}
\newcommand{\dx}{{d^3 x}}
\newcommand{\Deltatil}{\tilde{\Delta}}
\newcommand{\Dcal}{{\cal D}}
\newcommand{\DOS}{N}
\newcommand{\dos}{\nu}
\newcommand{\doss}{\nu_0}
\newcommand{\dosu}{\dos_{+}}
\newcommand{\dosd}{\dos_{-}}
\newcommand{\dosp}{\dos_{+}}
\newcommand{\dosm}{\dos_{-}}
\newcommand{\DOSV}{{N(0)}}
\newcommand{\DOSom}{{\DOS_\omega}}
\newcommand{\ef}{{\epsilon_F}}
\newcommand{\eF}{{\epsilon_F}}
\newcommand{\eft}{{\epsilon_F \tau}}
\newcommand{\eftoh}{\frac{\epsilon_F \tau}{\hbar}}
\newcommand{\eftauinv}{\frac{\hbar}{\epsilon_F \tau}}
\newcommand{\ekv}{\epsilon_{\kv}}
\newcommand{\ekvs}{\epsilon_{\kv\sigma}}
\newcommand{\ekvp}{\epsilon_{\kv'}}
\newcommand{\ekvps}{\epsilon_{\kv'\sigma}}
\newcommand{\elld}{{\ell_{\rm D}}}
\newcommand{\ells}{\ell_{\rm s}}
\newcommand{\Ev}{{\bm E}}
\newcommand{\ev}{{\bm e}}
\newcommand{\evth}{{\bm e}_{\theta}}
\newcommand{\evph}{{\bm e}_{\phi}}
\newcommand{\evs}{{\bm n}}%{{\bm e}_{\rm s}}
\newcommand{\evsz}{{\evs}_0}
\newcommand{\evsph}{(\evph\times\evz)}
\newcommand{\evx}{{\bm e}_{x}}
\newcommand{\evy}{{\bm e}_{y}}
\newcommand{\evz}{{\bm e}_{z}}
\newcommand{\fl}{{\eta}}
\newcommand{\fltil}{\tilde{\eta}}
\newcommand{\flitil}{\tilde{\fl_{\rm I}}}
\newcommand{\flrtil}{\tilde{\fl_{\rm R}}}
\newcommand{\fli}{{\fl_{\rm I}}}
\newcommand{\flr}{{\fl_{\rm R}}}
\newcommand{\fkv}{f_{\kv}}
\newcommand{\fkvs}{f_{\kv\sigma}}
\newcommand{\fo}{{f(\omega)}}
\newcommand{\fpo}{{f'(\omega)}}
\newcommand{\fomega}{{\omega}}
\newcommand{\fbeta}{f^{\beta}}
\newcommand{\fpin}{{f_{\rm pin}}}
\newcommand{\Fpin}{{F_{\rm pin}}}
\newcommand{\fe}{{f_{\rm e}}}
\newcommand{\fna}{{f_{\rm ref}}}
\newcommand{\Fe}{F}%_{\rm e}}
\newcommand{\Fev}{\Fv}%{\Fv_{\rm e}}
\newcommand{\Fbeta}{F^{\beta}}
\newcommand{\Fbetav}{\Fv^{\beta}}
\newcommand{\Fbetafactor}{\mu}
\newcommand{\Fhall}{F^{\rm Hall}}
\newcommand{\Fhallv}{\Fv^{\rm Hall}}
\newcommand{\Fren}{F^{\rm ren}}
\newcommand{\Frenv}{\Fv^{\rm ren}}
\newcommand{\Fvna}{\Fv^{\rm ref}}
\newcommand{\Fnav}{\Fv^{\rm ref}}
\newcommand{\Fna}{F^{\rm ref}}
\newcommand{\Fad}{\Fhall}
\newcommand{\Fzad}{F^{\rm (0)ad}}
\newcommand{\Fv}{{\bm F}}
\newcommand{\Fo}{F^{(1)}}
\newcommand{\Fw}{F_{\rm w}}%_{\rm e}}
\newcommand{\Fz}{F^{(0)}}
\newcommand{\Fzv}{\Fv^{(0)}}
\newcommand{\Fdelta}{\delta F}
\newcommand{\Fdeltav}{\delta \Fv}
\newcommand{\Fdel}{\delta \Fo}
\newcommand{\Fdelv}{\delta \Fv^{(1)}}
\newcommand{\Gv}{{\bm G}}
\newcommand{\gv}{{\bm g}}
\newcommand{\gr}{g^{\rm r}}
\newcommand{\gto}{g^{\rm t}}
\newcommand{\ga}{g^{\rm a}}
\newcommand{\Gr}{G^{\rm r}}
\newcommand{\Ga}{G^{\rm a}}
\newcommand{\Gto}{G^{\rm t}}
\newcommand{\Gat}{G^{\overline{\rm t}}}
\newcommand{\Gless}{G^{<}}
\newcommand{\Ggrt}{G^{>}}
\newcommand{\gless}{g^{<}}
\newcommand{\ggrt}{g^{>}}
\newcommand{\Gtil}{\tilde{G}}
\newcommand{\Gcal}{{\cal G}}
\newcommand{\gap}{\Delta_{\rm sw}}
\newcommand{\gammap}{\gamma_{+}}
\newcommand{\gammam}{\gamma_{-}}
\newcommand{\gammaz}{\gamma_{0}}
\newcommand{\grst}{|0\rangle}
\newcommand{\grsthc}{\langle0|}
\newcommand{\grft}{|\ \rangle}
\newcommand{\gyro}{\gamma}
\newcommand{\gyroz}{\gyro_{0}}
\newcommand{\hf}{\frac{1}{2}}
\newcommand{\HA}{{H_{A}}}
\newcommand{\HB}{H_{B}}
\newcommand{\Hv}{\bm{H}}
\newcommand{\He}{H_{\rm e}}
\newcommand{\Heff}{H_{\rm eff}}
\newcommand{\Hem}{H_{\rm em}}
\newcommand{\Hex}{H_{\rm ex}}
\newcommand{\Himp}{H_{\rm imp}}
\newcommand{\Hint}{H_{\rm int}}
\newcommand{\HR}{{H_{\rm R}}}
\newcommand{\Hs}{{H_{\rm S}}}
\newcommand{\Hsf}{{H_\spinflip}}
\newcommand{\Hso}{{H_{\rm so}}}
\newcommand{\Hsd}{H_{sd}}
\newcommand{\Hst}{{H_{\rm ST}}}
\newcommand{\Hw}{H_{\dw}}
\newcommand{\Hz}{H_{0}}
\newcommand{\hbarinv}{\frac{1}{\hbar}}
\renewcommand{\Im}{{\rm Im}}
\newcommand{\ime}{\gamma}%{\frac{1}{2\tau}}
\newcommand{\intinf}{\int_{-\infty}^{\infty}}
\newcommand{\intek}{\int_{-\ef}^{\infty}\! d\epsilon}
\newcommand{\intom}{\int\! \frac{d\omega}{2\pi}}
\newcommand{\intx}{\int\! {d^3x}}
\newcommand{\intk}{\int\! \frac{d^3k}{(2\pi)^3}}
\newcommand{\intr}{\int\! {d^3r}}
\newcommand{\intt}{\int_{-\infty}^{\infty}\! {dt}}
\newcommand{\ioh}{\frac{i}{\hbar}}
\newcommand{\iv}{\bm{i}}
\newcommand{\Ibar}{\overline{I}}
\newcommand{\Iv}{\bm{I}}
\newcommand{\Jex}{{J_{\rm ex}}}
\newcommand{\Jsd}{J_{sd}}
\newcommand{\Js}{{J_{\rm s}}}
\newcommand{\Jv}{\bm{J}}
\newcommand{\js}{j_{\rm s}}
\newcommand{\jsc}{{j_{\rm s}^{\rm c}}}
\newcommand{\jsv}{\bm{j}_{\rm s}}
\newcommand{\Jsv}{\bm{J}_{\rm S}}
\newcommand{\JSv}{\bm{J}_{\rm S}}
\newcommand{\JS}{J_{\rm S}}
\newcommand{\JStotv}{\bm{J}_{S,{\rm tot}}}
\newcommand{\jc}{j_{\rm c}}
\newcommand{\jci}{{{j}_{\rm c}^{\rm i}}}
\newcommand{\jce}{{{j}_{\rm c}^{\rm e}}}
\newcommand{\jatil}{{\tilde{j}_{\rm a}}}
\newcommand{\jctil}{{\tilde{j}_{\rm c}}}
\newcommand{\jcitil}{{\tilde{j}_{\rm c}^{\rm i}}}
\newcommand{\jcetil}{{\tilde{j}_{\rm c}^{\rm e}}}
\newcommand{\jstil}{{\tilde{j}_{\rm s}}}
\newcommand{\jtil}{{\tilde{j}}}
\newcommand{\jv}{\bm{j}}
\newcommand{\kB}{{k_B}}
\newcommand{\kb}{{k_B}}
\newcommand{\kv}{{\bm k}}
\newcommand{\kvxv}{\kv\cdot\xv}
\newcommand{\kvo}{{\kv_1}}
\newcommand{\kvp}{{\kv}'}
\newcommand{\kpq}{{k+\frac{q}{2}}}
\newcommand{\kmq}{{k-\frac{q}{2}}}
\newcommand{\kvpq}{{\kv}+\frac{\qv}{2}}
\newcommand{\kvmq}{{\kv}-\frac{\qv}{2}}
\newcommand{\kvopq}{{\kvo}+\frac{\qv}{2}}
\newcommand{\kvomq}{{\kvo}-\frac{\qv}{2}}
\newcommand{\kvppq}{{\kvp}+\frac{\qv}{2}}
\newcommand{\kvpmq}{{\kvp}-\frac{\qv}{2}}
\newcommand{\kf}{{k_F}}
\newcommand{\kF}{{k_F}}
\newcommand{\kfpm}{{k_{F\pm}}}
\newcommand{\kfmp}{{k_{F\mp}}}
\newcommand{\kfp}{k_{F+}}
\newcommand{\kfm}{k_{F-}}
\newcommand{\kfu}{k_{F+}}
\newcommand{\kfd}{k_{F-}}
\newcommand{\kfs}{k_{F\sigma}}
\newcommand{\kom}{{k_\omega}}
\newcommand{\Kp}{{K_\perp}}
\newcommand{\ktil}{\tilde{k}}
\newcommand{\lams}{{\lambda_{\rm s}}}
\newcommand{\lamv}{{\lambda_{\rm v}}}
\newcommand{\lamso}{{\lambda_{\rm so}}}
\newcommand{\lamz}{{\lambda_{0}}}
\newcommand{\Le}{{L_{\rm e}}}
\newcommand{\Lez}{{L_{\rm e}^0}}
\newcommand{\Leff}{L_{\rm eff}}
\newcommand{\Lb}{L_{\rm B}}
\newcommand{\Ldw}{L_{\dw}}
\newcommand{\Lsd}{{L_{sd}}}
\newcommand{\Ls}{{L_{\rm S}}}
\newcommand{\Lsw}{L_{\rm sw}}
\newcommand{\Lswdw}{L_{\rm sw-dw}}
\newcommand{\Linv}{{\frac{1}{L}}}
\newcommand{\lstil}{\tilde{l_\sigma}}
\newcommand{\Lv}{{\bm L}}
\newcommand{\mv}{{\bm m}}
\newcommand{\Mv}{{\bm M}}
\newcommand{\Mphi}{{M_{\phi}}}
\newcommand{\Mw}{{M_{\dw}}}
\newcommand{\Ms}{M_{\rm s}}
\newcommand{\MR}{{\rm MR}}
\newcommand{\Mz}{M_{0}}
\newcommand{\mus}{{g\mu_{B}}}
\newcommand{\mub}{\mu_B}
\newcommand{\muB}{\mu_B}
\newcommand{\muz}{\mu_0}
\newcommand{\muu}{\mu_+}
\newcommand{\mud}{\mu_-}
\newcommand{\muspin}{\mu_{\rm s}}
\newcommand{\nel}{n_{\rm e}}
\newcommand{\Ne}{N_{\rm e}}
\newcommand{\nv}{{\bm n}}
\newcommand{\Nimp}{N_{\rm imp}}
\newcommand{\nimp}{n_{\rm imp}}
\newcommand{\Ninv}{\frac{1}{N}}
\newcommand{\Nw}{N_{\dw}}
\newcommand{\nvortex}{n_{\rm v}}
\newcommand{\nvz}{{\nv}_0}
\newcommand{\nso}{n_{\so}}
\newcommand{\np}{{n}_+}
\newcommand{\nm}{{n}_-}
\newcommand{\nz}{{n}_0}
\newcommand{\om}{{\omega}}
\newcommand{\omegap}{\omega'}
\newcommand{\Omegatil}{\tilde{\Omega}}
\newcommand{\Omegap}{\Omega'}
\newcommand{\Omegapin}{\Omega_{\rm pin}}
\newcommand{\ompOm}{{\omega+\frac{\Omega}{2}}}
\newcommand{\ommOm}{{\omega-\frac{\Omega}{2}}}
\newcommand{\Omz}{\Omega_0}
\newcommand{\ompOmz}{{\omega+\frac{\Omz}{2}}}
\newcommand{\ommOmz}{{\omega-\frac{\Omz}{2}}}
\newcommand{\Omhf}{\frac{\Omega}{2}}
\newcommand{\phiz}{{\phi_0}}
\newcommand{\Phiv}{\bm{\Phi}}
\newcommand{\PhiB}{{\Phi_{\rm B}}}
\newcommand{\Ptil}{{\tilde{P}}}
\newcommand{\pv}{{\bm p}}
\newcommand{\Pv}{{\bm P}}
\newcommand{\PDOS}{P_{\DOS}}
\newcommand{\qv}{{\bm q}}
\newcommand{\qvxv}{\qv\cdot\xv}
\newcommand{\qtil}{{\tilde{q}}}
\newcommand{\ra}{\rightarrow}
\renewcommand{\Re}{{\rm Re}}
\newcommand{\rhow}{{\rho_{\dw}}}
\newcommand{\rhos}{{\rho_{\rm s}}}
\newcommand{\rhoS}{{\rho_{\rm s}}}
\newcommand{\rhoxy}{{\rho_{xy}}}
\newcommand{\RS}{{R_{\rm S}}}
\newcommand{\Rw}{{R_{\dw}}}
\newcommand{\rv}{{\bm r}}
\newcommand{\Rv}{{\bm R}}
\newcommand{\sd}{$s$-$d$}
\newcommand{\sigmav}{{\bm \sigma}}
\newcommand{\sigmaB}{\sigma_{\rm B}}
\newcommand{\sigmab}{\sigma_{\rm B}}
\newcommand{\sigmaz}{\sigma_0}
\newcommand{\sigmas}{\sigma_{\rm s}}
\newcommand{\se}{{s}}
\newcommand{\sev}{{\bm \se}}
\newcommand{\sevsf}{{\bm \se}_\spinflip}
\newcommand{\SE}{\Sigma}
\newcommand{\SEr}{\Sigma^{\rm r}}
\newcommand{\SEa}{\Sigma^{\rm a}}
\newcommand{\SEless}{\Sigma^{<}}
\newcommand{\sgn}{{\rm sgn}}
\newcommand{\sz}{{s}_0}
\newcommand{\sv}{{{\bm s}}}
\newcommand{\seth}{{\se}_\theta}
\newcommand{\seph}{{\se}_\phi}
\newcommand{\sez}{{\se}_z}
\newcommand{\so}{{\rm so}}
\newcommand{\spol}{{M}}
\newcommand{\spinflip}{{\rm sr}}
\newcommand{\svtil}{\tilde{\bm s}}
\newcommand{\stil}{\tilde{\se}}
\newcommand{\stilz}{\stil_{z}}
\newcommand{\stilpm}{\stil^{\pm}}
\newcommand{\stilpmz}{\stil^{\pm(0)}}
\newcommand{\stilpma}{\stil^{\pm(1{\rm a})}}
\newcommand{\stilpmb}{\stil^{\pm(1{\rm b})}}
\newcommand{\stilpmo}{\stil^{\pm(1)}}
\newcommand{\stilpara}{\stil_{\parallel}}
\newcommand{\stilperp}{\stil_{\perp}}
\newcommand{\Simpv}{{{\bm S}_{\rm imp}}}
\newcommand{\Simp}{{S_{\rm imp}}}
\newcommand{\Stot}{{S_{\rm tot}}}
\newcommand{\Stotv}{\bm{S}_{\rm tot}}
\newcommand{\Sh}{{\hat {S}}}
\newcommand{\Svh}{{\hat {\Sv}}}
\newcommand{\Sv}{{{\bm S}}}
\newcommand{\Svz}{{{\bm S}_0}}
\newcommand{\sumx}{{\int\! \frac{d^3x}{a^3}}}
\newcommand{\sumk}{{\sum_{k}}}
\newcommand{\sumkv}{{\sum_{\kv}}}
\newcommand{\sumqv}{{\sum_{\qv}}}
\newcommand{\sumom}{\int\!\frac{d\omega}{2\pi}}
\newcommand{\sumOm}{\int\!\frac{d\Omega}{2\pi}}
\newcommand{\sumomOm}{\int\!\frac{d\omega}{2\pi}\int\!\frac{d\Omega}{2\pi}}
\newcommand{\thickness}{{d}}
\newcommand{\thetaz}{{\theta_0}}
\newcommand{\tr}{{\rm tr}}
\newcommand{\To}{{\rm T}}
\newcommand{\Ta}{\overline{\rm T}}
\newcommand{\Tc}{{\rm T}_{C}}
\newcommand{\Tct}{{\rm T}_{\Ct}}
\newcommand{\tcmp}{\tau}
\newcommand{\tcmpi}{\tau_{\rm I}}
\newcommand{\tcmpinf}{\tau_{\infty}}
\newcommand{\tcmpz}{\tau_{0}}
\newcommand{\tcmpzp}{\tau_{0}'}
\newcommand{\Torqv}{{\bm \tau}}
\newcommand{\torque}{{\tau}}
\newcommand{\torquev}{{\bm \torque}}
\newcommand{\Torquev}{{\bm \tau}}
\newcommand{\torqueve}{\torquev}
\newcommand{\torquee}{\torque}
\newcommand{\torquew}{{\torque_{\dw}}}
\newcommand{\tautil}{{\tilde{\tau}}}
\newcommand{\taup}{\tau_{+}}
\newcommand{\taum}{\tau_{-}}
\newcommand{\tauw}{\tau_{\dw}}
\newcommand{\taus}{\tau_{\rm s}}
\newcommand{\tausf}{\tau_\spinflip}
\newcommand{\thetast}{\theta_{\rm st}}
\newcommand{\ttil}{{\tilde{t}}}
\newcommand{\tinf}{t_\infty}
\newcommand{\tz}{t_0}
\newcommand{\Ubar}{\overline{U}}
\newcommand{\Ueff}{U_{\rm eff}}
\newcommand{\Uz}{U_0}
\newcommand{\Uv}{U_V}
\newcommand{\vc}{{v_{\rm c}}}
\newcommand{\ve}{{v_{\rm e}}}
\newcommand{\vev}{{\vv_{\rm e}}}
\newcommand{\vv}{\bm{v}}
\newcommand{\vs}{{v_{\rm s}}}
\newcommand{\vsv}{{\vv_{\rm s}}}
\newcommand{\vw}{v_{\rm w}}
\newcommand{\vf}{{v_F}}
\newcommand{\vimp}{v_{\rm imp}}
\newcommand{\vi}{{v_{\rm i}}}
\newcommand{\Vi}{{V_{\rm i}}}
\newcommand{\Vso}{v_{\rm so}}
\newcommand{\vtil}{{{v_0}}}
\newcommand{\Vpin}{{V}_{\rm pin}}
\newcommand{\Vinv}{\frac{1}{V}}
\newcommand{\vz}{{v_0}}
\newcommand{\Vz}{{V_0}}
\newcommand{\Vcal}{{\cal V}}
\newcommand{\Vztil}{{\tilde{V_0}}}
\newcommand{\Ws}{{W_{\rm S}}}
\newcommand{\Xtil}{{\tilde{X}}}
\newcommand{\xv}{{\bm x}}
\newcommand{\Xv}{{\bm X}}
\newcommand{\xvp}{{\bm x}_{\perp}}
\newcommand{\xw}{{z}}
\newcommand{\Xz}{{X_0}}
\newcommand{\Zs}{Z_{\rm S}}
\newcommand{\Zz}{Z_{0}}
\newcommand{\ztil}{u}%{\tilde{\xw}}
\newcommand{\zh}{\hat{z}}

\renewcommand{\nimp}{n_{\rm i}}
\renewcommand{\vimp}{v_{\rm i}}

%%%%%%%%%%%%%%%%%%%%%%%%%%%%%%%

\section{Introduction}

Spintronics aims at using the information carried by the electrons' spin in solids.
For this purpose, establishing reliable methods to create, transfer and detect the spin current is an urgent task.
Compared to the charge transports, spin transports have one serious fundamental difficulty.
That is the non-conservation of the spin in solids.
This limits the range of the spin transmission to be less than the spin diffusion length, which is typically $\mu$m scale in metals.

The non-conservation of the spins is expressed by a source term in the continuity equation for the spin 
\begin{align}
\dot{s}^\alpha+\nabla \cdot\jsv^\alpha={\cal T}^\alpha.
  \label{sconteq}
\end{align}
Here $s$ and $\jsv$ are the spin density and spin current, respectively, $\alpha=x,y,z$ is the spin direction and ${\cal T}$ is the spin relaxation torque resulting in the non-conservation of the spin.
In the most cases in metals, the dominant origin of ${\cal T}$ is the spin-orbit interaction.

Although the relaxation torque term is essential in spin transports, it has so far been treated only on the phenomenological ground.
The continuity equation is equivalent to the Boltzmann equation, which is useful in discussing the spin transports.
The Boltzmann equation for the distribution function of each spin channel was discussed by Son et al. \cite{Son87} and later by Valet and Fert \cite{Valet93} in the context of the giant magnetoresistance in multilayer systems.
In their analysis, they approximated the spin relaxation torque as proportional to the inverse of a spin relaxation time $\tau_{\rm sf}$ 
and to some unknown function representing a driving force for the spin accumulation.
The driving force was written in terms of what they called the spin chemical potential $\muspin$.
The relaxation torque was approximated as ${\cal T} ^z =\muspin/\tau_{\rm sf}$.
They argued that $\muspin$ satisfies the diffusion equation,
$\nabla^2 \muspin =-\ell_{\rm sf}^{-2} \muspin$,  with the diffusion length $\ell_{\rm sf}\propto \sqrt{\tau_{\rm sf}}$.
Microscopic calculation for $\muspin$ has not been done so far.

The diffusion equation for the spin has been widely used to discuss recent spin transports in metallic junctions \cite{Takahashi08}.
The decay of spin transport has been confirmed in non-local spin injection experiments \cite{Jedema01,Kimura07,Seki08}, which
 indicate that the spin diffusion decays with a decay length of 350-500nm in Cu and  100nm in Au at room temperature.
Although the spin diffusion equation appears to be so far successful, the phenomenological treatment of the spin relaxation term and the spin chemical potential must be improved 
to consider the spin transport seriously.

Besides  diffusive spin current, there is another spin current that is driven by an effective field.
In contrast to the diffusive one, this field-driven contribution  
should not decay in uniform (single domain) ferromagnets, since the ratio of the spin current and the charge current is determined by the spin polarization ratio of the material, which is a statistical mechanical quantity. 
The field-driven (local) spin current and the diffusive spin current behave differently, as was recently  demonstrated theoretically  in the case of the inverse spin Hall effect \cite{Takeuchi10}.

In the field of the current-driven magnetization dynamics,  the spin relaxation torque has been studied from the microscopic viewpoint \cite{KTS06,TKS_PR08,TE08}.
In this context, Eq. (\ref{sconteq}) gives the expression for the torque acting on the spin density $\sv$ as 
\begin{align}
\tau^\alpha = -\nabla \cdot\jsv^\alpha+{\cal T}^\alpha.
\end{align}
In the adiabatic limit, i.e., slowly varying magnetization, and under uniform current, the first term reduces to 
$\nabla \cdot\jsv^\alpha=(P/2e)(\jv\cdot\nabla)\sv^\alpha$, where $P$ is the spin polarization of the current \cite{TKS_PR08,TE08}, namely to the adiabatic spin-transfer torque.
When the spin-relaxation sets in, the conduction electron no longer  follows the magnetization profile, and new contribution to the torque arises from the ${\cal T}$ term.
This torque was shown to be
\begin{align}
{\cal T} =
-\beta \frac{P}{es^2} 
(\sv\times (\jv\cdot\nabla)\sv),  \label{betatorque}
\end{align}
where $\beta$ is a coefficient inversely proportional to the spin relaxation time $\tau_{\rm s}$ \cite{KTS06,TE08}.
This torque, called $\beta$ term, turned out to be essential in determining the efficiency of the current-driven domain wall motion \cite{TK04,Zhang04,Thiaville05}.
The magnitude of the parameter $\beta$ has recently been intensively studied experimentally by measuring the domain wall speed under current \cite{Thomas06,Moore09}.
Theoretical formulation for estimating $\beta$ in the first-principles calculations was carried out recently \cite{Garate09}.

The spin relaxation torque has been studied also from the viewpoint of how to define the spin current.
It was discussed that the spin relaxation torque contains a term written as a divergence of the torque dipole density, $\Pv$ \cite{Culcer04}.
Generalized argument was given by Shi et al. \cite{Shi06}, where they discussed that the $z$ component of the relaxation torque is written as a divergence, 
\begin{align}
{\cal T}^z= -\nabla\cdot\Pv, 
\end{align}
if the system has the inversion symmetry. 
This means that the total torque integrated over the system should vanish.
Shi et al. also argued that if the relaxation torque is a divergence of $\Pv$, one can define a spin current that is conserved. 
In fact, defining 
$\tilde{\jsv}\equiv \jsv+\Pv$, the continuity equation (\ref{sconteq}) reduces to $\dot{s}^z+\nabla\cdot \tilde{\jsv}=0$.
The explicit form of the torque dipole density was not calculated in Ref. \cite{Shi06}.
Obviously, in the presence of the inhomogeneity of the magnetization, the $\beta$ torque (\Eqref{betatorque}) cannot be written as a divergence, and thus it indeed represents the spin angular momentum lost by the spin relaxation.

The result that the spin relaxation torque is given by a derivative of the applied electric field $\Ev$ is understood as follows.
The spin relaxation torque should of course vanish when $\Ev=0$.
It cannot be directly proportional to $\Ev$, since field-driven spin current in uniform ferromagnets should not decay. 
Therefore, the simplest expression for the relaxation torque is a derivative of the field.
It is not, however, obvious whether it should be always written in a rotationally invariant way, or if it can be
anisotropic, since the rotational invariance is broken in uniform ferromagnets because of the magnetization.

The first aim of the present paper is to calculate the spin relaxation torque microscopically in the presence of the applied electric field. 
The spin relaxation mechanism we take into account is the spin-orbit interaction  due to random impurities. 
Our explicit calculation reveals that the spin relaxation torque is not always rotationally symmetric in uniform ferromagnets, but is generally given by
\begin{align}
{\cal T}^z = \gamma (\nabla\cdot \Ev)
+\delta\gamma(\partial_z E_z),
\label{TauE}
\end{align}
where $z$ axis is along the magnetization, 
 $\gamma$ and  $\delta\gamma$ are coefficient proportional to the inverse spin relaxation time.
The spin relaxation torque is, therefore, anisotropic.
Relaxation torque of \Eqref{TauE} indicates that the torque dipole density is given by
\begin{align}
\Pv = -(\gamma  \Ev +\delta\gamma (\nv\cdot \Ev)\nv),
\end{align}
where $\nv$ represents the direction of the magnetization.
These anisotropic behaviors of the transport quantities is common when spin-orbit interaction exists, as is well-known in charge transport as the anisotropic magnetoresistance (AMR) \cite{Mcguire75}. 

The second aim of the paper is to study the spin current on the same microscopic footing as the relaxation torque.
We show that the spin current is made up of a field-driven contribution which is local and the diffusive one with nonlocality.
The field-driven contribution is anisotropic like the AMR effect for the charge current 
(we call the effect as spin AMR effect).
The diffusive contribution is given as a gradient of a spin chemical potential, $\muspin$.
We will derive the linear-response expression for the spin chemical potential. 
Our microscopic study on the spin current demonstrates the validity of the half-phenomenological treatments \cite{Valet93}.

\section{Model}

We consider the conduction electron system taking account of the spin-orbit interaction, the impurity scattering without spin flip, and the applied electric field. 
The Hamiltonian of the system is given as
$H=H_0+ \Hso+\Hem+\Himp$, where 
$H_0$
 is the free electron Hamiltonian including the uniform magnetization, $\Hso$ is the spin-orbit interaction, $\Hem$ is the interaction with the gauge field representing the applied electric field, and $\Himp$ is the spin-independent impurity scattering.
The free part reads 
\begin{align}
H_0\equiv \sum_{\kv\sigma}\epsilon_{\kv\sigma} c^\dagger_{\kv\sigma} c_{\kv\sigma} ,
\end{align}
 where 
the electron creation and annihilation operators are denoted by $\cdag$ and $c$, respectively, $\epsilon_{\kv\sigma}\equiv \frac{k^2}{2m}-\ef-\sigma \spol$, $\ef$ is the Fermi energy, $\spol$ is the spin splitting due to the magnetization and $\sigma\equiv \pm$ represents the spin.
The spin-orbit interaction is represented by
$\Hso=\Hso^0+\Hso^{A}$,
 where 
($\sigma_k$ ($k=x,y,z$) is the Pauli matrix)
\begin{align}
\Hso^0 &=  -\frac{i}{2}
\sum_{ijk} \epsilon_{ijk}\intx  (\nabla_i \Vso^{(k)}) 
(\cdag \vvec{\nabla}_j \sigma_k c)
\\
\Hso^{A} &= - e\sum_{ijk} \epsilon_{ijk}\intx  
 (\nabla_i \Vso^{(k)}) A_j(\xv,t) (\cdag \sigma_k c).
\end{align}
(We suppress the spin index when obvious,  namely, $c=(c_+,c_-)$. )
The spin-orbit potential $\Vso^{(k)}$ is assumed to arise from random impurities and to depend on the spin direction ($k$).
The averaging over the spin-orbit potential is carried out as 
\begin{align}
\average{ \Vso^{(k)}(\pv)  \Vso^{(\gamma)}(-\pv')}_{\rm i} =\nso\lamso^2 \delta_{\pv,\pv'}\delta_{k\gamma}, 
\label{vsav}
\end{align} 
where $\nso$ and $\lamso$ are the concentration of the spin-orbit impurities and the strength of the interaction, respectively.
The average of the spin-orbit potential at the linear order is zero in our model, and thus we do not take account of the the anomalous Hall and spin Hall effects. 
We consider a case where electric field 
$\Ev(\xv,t)(\equiv -\dot{\Av}(\xv,t))$ is position and time dependent. 
The electromagnetic interaction is written as
\begin{align}
\Hem = -\frac{e}{m} \sum_{\kv,\qv}
\sum_i k_i A_i(\qv,\Omega) (\cdag_{\kvmq}c_{\kvpq}),
\end{align}
where $\Omega$ is the frequency of the electric field.
We will consider the limit of small $\Omega$ and small $q$.
The scattering by the normal impurities is represented by
\begin{align}
\Himp &= 
\sum_{i=1}^{\Nimp}\sum_{\kv\kv'} 
\frac{\vimp}{N} e^{i(\kv-\kv')\cdot\Rv_i} \cdag_{\kv'}c_{\kv},
\end{align}
where $\vimp$ represents the strength of the impurity potential, $\Rv_i$ represents the position of random impurities, $\Nimp$ is the number of impurities, and 
$N\equiv V/a^3$ is number of sites.
To estimate physical quantities, we take the random average over impurity positions in a standard manner \cite{TKS_PR08}.

To derive the spin continuity equation, \Eqref{sconteq}, we derive the equation of motion for the spin density, 
$\sev(\xv,t)\equiv \average{\cdag(\xv,t)\sigmav c(\xv,t)}$
($\average{\ }$ represents the quantum average).
The time development of the spin density reads 
\begin{align}
\dot{s}^\alpha &= 
i\average { [H,c^\dagger]\sigma^\alpha c+ c^\dagger\sigma^\alpha [H,c] },\label{eqcommutators}
\end{align}
where $H$ is the total Hamiltonian of the system.
The commutators are calculated as in Appendix \ref{APP:eqofmo}, and \Eqref{eqcommutators}
turns out to be \Eqref{sconteq}, namely 
\begin{align}
\dot{s}^\alpha = -\nabla \cdot\jsv^\alpha+{\cal T}^\alpha, \nonumber
\end{align}
 with the spin current given as
\begin{align}
\js^{\alpha} \equiv \js^{{\rm (n)},\alpha}+\js^{{\rm so},\alpha}
, 
\end{align}
where 
\begin{align}
 \jspin{,i}{{\rm (n)},\alpha} 
  &\equiv -\frac{i}{2m} 
\average{ \cdag \sigma^\alpha \vvec{\nabla}_i c }
 -\frac{e}{m}A_i 
\average{ \cdag \sigma_\alpha c }
\nnr
&\equiv 
 \jspin{,i}{(0),\alpha} + \jspin{,i}{A,\alpha}  ,
\label{spincurrentsdefs}
\end{align}
and 
\begin{align}
 \jspin{,i}{\so, \alpha} 
  \equiv 
-\sum_{j} \epsilon_{ij\alpha} 
(\nabla_j\Vso^{(\alpha)}) \average{ \cdag c}. \label{jssodef}
\end{align}
The relaxation torque reads 
\begin{align}
{\cal T}^{\alpha}
\equiv {\cal T}_{\so}^{\alpha} +{\cal T}_{\so}^{A,\alpha}, \label{totaltaudef}
\end{align}
 where 
\begin{align}
{\cal T}_{\so}^{\alpha} 
&\equiv 
i \sum_{ijkl} \epsilon_{ijk} \epsilon_{\alpha lk} (\nabla_i\Vso^{(k)})
\average{ \cdag \sigma_l \vvec{\nabla}_j c},
\label{Tsodef} 
\\
{\cal T}_{\so}^{A,\alpha} 
& \equiv 
2e \sum_{ijkl} \epsilon_{ijk} \epsilon_{\alpha lk} (\nabla_i\Vso ^{(k)} ) A_j
\average{ \cdag \sigma_l c}.\label{TsoAdef}
\end{align}

The spin relaxation torque depends on the definition of the spin current. 
For instance, if we redefine the spin current as
 ${\jsv' } ^\alpha \equiv {\jsv^\alpha }-\Cv^\alpha$, where $\Cv$ is a vector, the continuity equation (\ref  {sconteq}) becomes 
$\dot{s}^\alpha+\nabla \cdot{ \jsv' } ^\alpha={{\cal T}'} ^\alpha$,  where 
the relaxation torque reads ${{\cal T}'} ^\alpha\equiv {{\cal T}} ^\alpha+\nabla\cdot \Cv^\alpha$.
This ambiguity of spin current definition of course does not affect physical quantities such as the total torque acting on the spin density, which is given by $\dot{\sv}$.

\section{Spin relaxation torque}

We calculate the spin relaxation torque as a linear response to the applied electric field. 
The uniform magnetization is chosen as along $z$ axis.
The spin-orbit interaction is included to the second order.
%
%%%%%%%%%%%
\begin{figure}[tbh]
\begin{center}
\includegraphics[width=0.45\hsize]{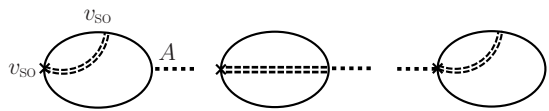}
\caption{ 
Feynman diagrams representing the relaxation torque.
%The last diagram rises from $\Hso^{A}$.
Solid lines represent the electron Green's function with the lifetime  ($\tau_\sigma$) included, 
$\Vso$ represents the spin-orbit interaction (double dashed line)  and dotted line represents the interaction with the gauge field ($A$).
The first two diagrams are the contributions to ${\cal T}_{\so}^{\alpha}$
and the last diagram is the contribution to ${\cal T}_{\so}^{A,\alpha}$.
The vertex marked by cross represents the relaxation torque.
\label{FIGspinrelxationdiag1}
}
\end{center}
\end{figure}
%%%%%%%%%%%%%%
%
The contributions to the relaxation torque, \Eqref{totaltaudef}, are shown in Fig. \ref{FIGspinrelxationdiag1}.
The leading contribution for small $1/(\ef\tau)$ and $q\ell$ 
($\ell$ is the electron mean free path)
turns out to be the first diagram in Fig. \ref{FIGspinrelxationdiag1}, which reads 
(see Appendix \ref{SEC:appA} for details)
\begin{align}
{\cal T}^{\alpha} 
&= \delta_{\alpha,z}
\frac{2}{45\pi}
\nso \lamso^2 \frac{e}{m^2}
%\sumOm\sum_{\qv} e^{-i\qv\cdot\xv} e^{i\Omega t} \Omega  (3\qv\cdot\Av +q_z A_z)
(3\nabla\cdot\dot{\Av}+\nabla_z\dot{A}_z)
\nnr
&\times 
\sum_{\kv\kv'}
k^2 (k')^4 
\sum_{\sigma=\pm} \sigma  \green{\kv,\sigma}{\ret} 
\green{\kv',-\sigma}{\ret} (\green{\kv',-\sigma}{\adv})^2 
+{\rm c.c.},
\label{Tauzdominant}
\end{align}
where $\green{\kv\sigma}{\ret}$ and $\green{\kv\sigma}{\adv}$ are the retarded and advanced electron Green's functions, respectively, carrying the wave vector $\kv$ and spin $\sigma$ with zero frequency. 
As we see, only $z$ component of the torque is finite.
The Green's functions include the lifetime arising from the self-energy process due to normal impurities and the spin-orbit interaction.
The inverse lifetime for the electron with spin $\sigma(=\pm)$ is given as
\begin{align}
{\tau_\sigma}^{-1}
=
2\pi \nimp\vimp^2 \dos_\sigma
(1+\kappa_{z,\sigma}+\kappa_{\perp} \gamma_\sigma),
\end{align}
where $\dos_\sigma$ is the spin-resolved electron density of states, $\nimp$ and $\vimp$ are the concentration and the potential strength of the impurities,
$\kappa_{z,\sigma}\equiv \frac{1}{3}\frac{\nso\lamso^2}{\nimp\vimp^2}k_{F\sigma}^4 $
and  
$\kappa_{\perp} \equiv \frac{2}{3}\frac{\nso\lamso^2}{\nimp\vimp^2}k_{F+}^2 k_{F-}^2 $
are dimensionless ratios of the spin-orbit interaction to the normal impurity scattering
($\kfs$ is the spin-dependent Fermi wavelength), and 
$\gamma_\sigma\equiv\frac{\dos_{-\sigma}}{\dos_\sigma}$.
The total relaxation torque  is therefore given by (\Eqref{TauE}),
\begin{align}
{\cal T}^z = \gamma (\nabla\cdot \Ev)
+\delta\gamma(\partial_z E_z),
\end{align}
where  
\begin{align}
\gamma
&\equiv
 \frac{8\pi e}{15m^2}
\nso \lamso^2 
\dosu \dosd \kfu^2 \kfd^2 
\sum_{\sigma=\pm} \sigma \kfs^2\tau_{\sigma}^2
%(\kfu^2\tau_+^2-\kfd^2\tau_-^2), \label{gammaEres}
\end{align}
and $\delta\gamma\equiv \gamma/3$.
The parameter $\gamma$ is proportional to the spin flip rate due to the spin-orbit interaction.
%(It vanishes when the spin splitting is zero as is obvious.)
Our result indicates that the 
relaxation torque is zero in uniform ferromagnet when uniform electric field is applied.
Thus the spin current does not decay in this case.
In fact,  the static solution of \Eqref{sconteq} with ${\cal T}^z=0$ is 
$\js^z={\rm constant}$. 

The degree of the asymmetry, $\delta\gamma/\gamma$, is not universal but is model dependent.
For instance, in the case of junction with weak electron hopping at point-like leads, $\delta \gamma$ vanishes \cite{Takezoe10}.

\section{Spin current}

\begin{figure}[tbh]
\begin{center}
\includegraphics[width=0.6\hsize]{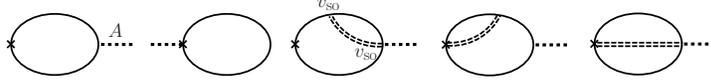}
\caption{ 
Feynman diagrams representing the local contribution to the spin current
(the first two terms in \Eqref{jsexpression2}).
The first three diagrams correspond to $ \jspin{}{{\rm (n)}}$, and the last two diagrams represent the contribution from the anomalous velocity,
 $\jspin{}{\so}$.
Dotted and double dashed lines denote the interaction with the applied electric field and the spin-orbit interaction, respectively.
The vertex marked by cross represents the spin current.
\label{FIGspincurrent}
}
\end{center}
\end{figure}

\begin{figure}[tbh]
\begin{center}
\includegraphics[width=0.65\hsize]{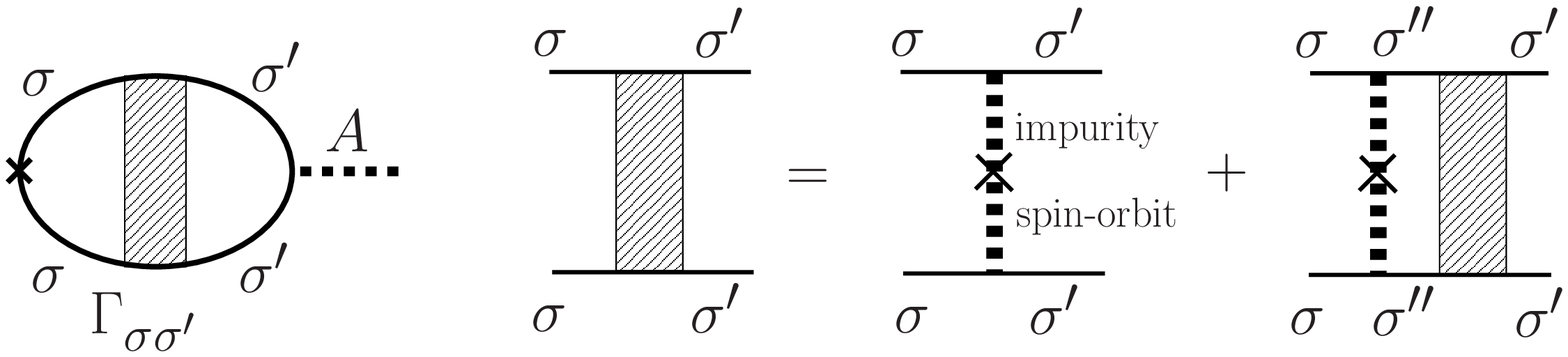}
\caption{ 
Left: The vertex correction contribution to $\js^{{\rm (n)},z}$,
resulting in the diffusive spin current
(the last term of \Eqref{jsexpression2}).
Right:
$\Gamma_{\sigma\sigma'}$, which is a ladder process of the successive electron scattering by the normal impurity and the spin-orbit interaction (represented by thick dotted lines)
connecting the  spin indices $\sigma$ and $\sigma'$.
\label{FIGdiffusion}
}
\end{center}
\end{figure}

We here calculate the spin current within the same formalism.
Within the linear response theory, the spin current is calculated by estimating the Feynman diagrams shown in Figs. \ref{FIGspincurrent} and \ref{FIGdiffusion}, which correspond to the field-induced contribution and the effect of the diffusive electron motion, respectively.

%%%%%%%%%%%
%%%%%%%%%%%%%%
%
%%%%%%%%%%%
%%%%%%%%%%%%%%
%

We first estimate the normal part of the spin current, 
$\jspin{,i}{{\rm (n)},z}$ (\Eqref{spincurrentsdefs}), shown in the first two diagrams in Fig. \ref{FIGspincurrent}.
% \equiv \jspin{,i}{(0),z} +\jspin{,i}{A,z} $.
The contribution $\jspin{,i}{{\rm (n)},z}$ is defined including the anomalous contribution from the electromagnetic gauge field, $\jspin{,i}{A,z}$.  
Its dominant contribution in the limit of small $\Omega$ and small $q$  is calculated as 
(see Appendix \ref{SEC:appB} for detail)
\begin{align}
\jspin{,i}{{\rm (n)},z} 
&=
 -i\frac{1}{2\pi} \frac{e^2}{m^2}
\sum_{\kv\qv}e^{-i\qv\cdot\xv}\sumOm e^{i\Omega t}
\sum_{j}
 \Omega A_j(\qv,\Omega) 
\sum_{\sigma} \sigma \lt[
k_i k_j\green{\kvmq,\sigma}{\ret} \green{\kvpq,\sigma}{\adv} 
\rt.
\nnr
&
\lt. +
\sum_{\kv'\sigma'}
k_i k'_j
\green{\kvmq,\sigma}{\ret} \green{\kvpq,\sigma}{\adv} 
\green{\kvpmq,\sigma'}{\ret} \green{\kvppq,\sigma'}{\adv} 
\nimp\vimp^2 \Gamma_{\sigma\sigma'}(\qv,\Omega)
\rt]
\label{spincurrent1}
\end{align}
In \Eqref{spincurrent1}, the first term is the contribution shown in the left of Fig. \ref{FIGspincurrent},
and the second term is the contribution from the vertex correction (Fig. \ref{FIGdiffusion}).

The factor $\Gamma_{\sigma\sigma'}(\qv,\Omega)$ contains all the 
vertex corrections due to the normal impurities and the spin-orbit interaction shown in Fig. \ref{FIGdiffusion}.
The equation of motion for $\Gamma_{\sigma\sigma'}$ is derived in the same manner as in Ref. \cite{Hikami80} carried out in the context of quantum correction  (the diffusion without the spin-orbit interaction was considered in Ref. \cite{Ban09}). 
The equation is obtained as
\begin{align}
\Gamma_{\sigma\sigma} 
&= (1+\kappa_z)(1+\Pi_\sigma\Gamma_{\sigma\sigma})
+\kappa_\perp \Pi_{-\sigma}\Gamma_{-\sigma,\sigma}
\nnr
\Gamma_{\sigma,-\sigma} 
&= \kappa_\perp(1+\Pi_{-\sigma}\Gamma_{-\sigma,-\sigma})
+(1+\kappa_z)  \Pi_{\sigma}\Gamma_{\sigma,-\sigma},
\label{Gammaeqs}
\end{align}
where 
\begin{align}
\Pi_\sigma(\qv,\Omega)
&\equiv \nimp\vimp^2 \sumkv 
 \green{\kvmq,\sigma}{\ret} \green{\kvpq,\sigma}{\adv} \nnr
&\simeq 
 [1-(D_\sigma q^2 \tau_\sigma+\kappa_z+\kappa_\perp\gamma_\sigma)],
\label{Pires}
\end{align}
where $D_\sigma \equiv \frac{(\kfs)^2}{3m^2}\tau_\sigma$ is the diffusion constant.
Neglecting quantities of order of $(\kappa_z,\kappa_\perp)^2$, \Eqref{Gammaeqs} is solved as
\begin{align}
\Gamma_{\sigma\sigma} 
&= \frac{1+\kappa_z-(1+2\kappa_z)\Pi_{-\sigma}}
{[1-(1+\kappa_z)\Pi_+][1-(1+\kappa_z)\Pi_-]}
\nnr
\Gamma_{\sigma,-\sigma} 
&= \frac{\kappa_\perp}
{[1-(1+\kappa_z)\Pi_+][1-(1+\kappa_z)\Pi_-]}.
\label{Gammaeqs2}
\end{align}
Using \Eqref{Pires}, we obtain $\Gamma_{\sigma\sigma'}$  as  (assuming the rotational symmetry for the wave vectors when averaging over the spin-orbit potential)
\begin{align}
\Gamma_{\sigma\sigma} 
&= \frac{1}
{D_\sigma q^2 \tau_\sigma+\kappa_\perp\gamma_\sigma}
\nnr
\Gamma_{\sigma,-\sigma} 
&= \frac{\kappa_\perp}
{[D_+ q^2 \tau_++\kappa_\perp\gamma_+]
 [D_- q^2 \tau_-+\kappa_\perp\gamma_-]}.
\label{Gammaeqs3}
\end{align}
By use of \Eqref{Gammaeqs3} and summing over the wave vectors,
the normal spin current, \Eqref{spincurrent1}, reads
\begin{align}
\jspin{,i}{{\rm (n)},z} 
&=
\sigma_{{\rm s}}^0 E_i
- \nabla_i \muspin,
\label{jsexpression0}
\end{align}
where 
$\sigma_{{\rm s}}^0 \equiv e \sum_{\pm} (\pm) D_\pm \dos_\pm$ is the bare spin conductivity divided by $e$.
The first term of \Eqref{jsexpression0} is the field-driven contribution.
The second gradient term is a diffusive contribution (vertex corrections), arising from the spin accumulation.
The effective potential describing the spin accumulation, $\muspin$, 
reads
\begin{align}
\muspin \equiv
\intx' \chi(\xv-\xv') (\nabla\cdot \Ev)(\xv'), \label{muspin}
\end{align}
where $\chi$ is a correlation function arising from the electron diffusion, given as ($V$ is the system volume)
\begin{align}
\chi(\xv)
&\equiv 
- \sum_{\pm}  (\pm)  \sigma_{\pm}
\frac{1}{V}\sumqv \frac{e^{-i\qv\cdot\xv}}
{q^2+(\ls{,\pm})^{-2}}.   \label{chidef}
\end{align}
Here
$\sigma_\pm \equiv e D_\pm \dos_\pm$ is the spin-resolved Boltzmann conductivity divided by $e$, and the correlation length is given as
\begin{align}
\ls{,\sigma} 
%=\sqrt{\frac{D_\sigma\tau_\sigma}{\kappa_\perp \gamma_\sigma}}
= \sqrt{D_\sigma\tau_{{\rm s},\sigma}}.
\end{align}
The lifetime of the spin $\sigma$ electron reads 
$\tau_{{\rm s},\sigma}
\equiv {\tau_\sigma}/{(\kappa_{\perp} \gamma_\sigma)}$.
Defining $\muspin=\mu_+-\mu_-$, 
we see that spin-resolved effective potential satisfies
\begin{align}
(-\nabla^2 + (\ls{,\sigma})^{-2})
\mu_\sigma 
= -\sigma_\sigma (\nabla\cdot\Ev)
.\label{mudiffusioneq}
\end{align}

In three-dimensions, the correlation function reads 
\begin{align}
\chi(\xv)=\frac{1}{4\pi |\xv|}\sum_{\pm} (\pm){\sigma_{\pm}}e^{-|\xv|/\ls{,\pm}}.
\end{align}

The local part of the spin current arises also from the anomalous spin current due to the spin-orbit interaction, defined in \Eqref{jssodef}.
This contribution is calculated by evaluating the last two diagrams in Fig. \ref{FIGspincurrent}
as (see Appendix \ref{SEC:appB})
\begin{align}
\jspin{,i}{\so,z} 
&=
\delta \sigma_{{\rm s}} (1-\delta_{i,z}) E_i,
\label{jssores2}
\end{align}
where
\begin{align}
\delta \sigma_{{\rm s}} 
&\equiv 
\frac{\pi}{9}
\nso\lamso^2 \frac{e}{m}  
\sum_{\pm}(\pm) (k_{F\pm})^4 (\dos_{\pm})^2 \tau_\pm .
%\nnr & = \frac{2\pi^2}{3a^3}
%\nso\lamso^2 \sum_{\pm}(\pm) \sigma_{\pm} k_{F\pm} (\dos_{\pm})^2  .
\label{jssores3}
\end{align}
This spin-orbit correction to the spin conductivity is anisotropic, resulting in a spin version of the anisotropic magnetoresistance (AMR) effect, namely, spin AMR effect.

From Eqs. (\ref{spincurrent1})(\ref{Gammaeqs3})(\ref{jssores2}), the leading contribution to the spin current for small $\qv$ and $\Omega$ is obtained as the sum of the local part driven by the electric field and the diffusive part as
%\begin{align}
%\jspin{,i}{z}=\sigma_{{\rm s}} E_i -\delta\sigma_{\rm s}E_z\delta_{i,z}- \nabla_i \muspin , \label{jsexpression2}
%\end{align}
\begin{align}
\jv_{{\rm s}}^{z}=\sigma_{{\rm s}} \Ev -\delta\sigma_{\rm s} (\nv\cdot\Ev) \nv 
- \nabla \muspin , 
\label{jsexpression2}
\end{align}
where 
$\sigma_{{\rm s}} \equiv \sigma_{{\rm s}}^{0}   +\delta \sigma_{{\rm s}} $ and $\nv$ is the unit vector along the magnetization.
In terms of the angle $\theta$ defined by
$\cos\theta\equiv (\nv\cdot\Ev)/E$, the magnitude of the field-driven (local) current reads
\begin{align}
j_{{\rm s}}^{{\rm loc},z}=\sqrt{ (\sigma_{{\rm s}\parallel})^2 
 + ((\sigma_{{\rm s}\perp })^2-(\sigma_{{\rm s}\parallel})^2) \sin^2\theta},
\label{jsmag}
\end{align}
where $\sigma_{{\rm s}\parallel}\equiv \sigma_{{\rm s}}^{0}$ and $\sigma_{{\rm s}\perp}\equiv \sigma_{{\rm s}}$.
When the degree of the anisotropy is small, the spin current becomes 
\begin{align}
\frac{j_{{\rm s}}^{{\rm loc},z}}{E}
=\sigma_{{\rm s}\parallel}\lt(1+ 
 \hf\lt( \frac{\sigma_{{\rm s}\perp }}{\sigma_{{\rm s}\parallel}} \rt)^2 \sin^2\theta \rt).
\label{jsmag2}
\end{align}
We define the magnitude of the spin AMR as 
\begin{align}
\frac{\Delta\rho_{{\rm s}}}{\rho_{{\rm s}\perp}}
& \equiv \frac{\rho_{{\rm s}\parallel}-\rho_{{\rm s}\perp}} {\rho_{{\rm s}\perp}} 
= \frac{ \delta \sigma_{{\rm s}} / \sigma_{{\rm s}} }{1-\delta \sigma_{{\rm s}} / \sigma_{{\rm s}}},
\end{align}
where $\rho_{{\rm s}\alpha} \equiv (\sigma_{{\rm s}\alpha})^{-1}$ ($\alpha=\parallel,\perp$).

\section{Spin injection}

\begin{figure}[tbh]
\begin{center}
\includegraphics[width=0.5\hsize]{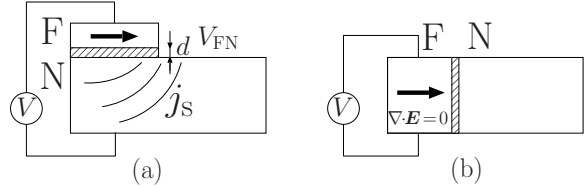}
\caption{ 
(a) Creation of diffusive spin current (spin injection) by applying 
the electric voltage perpendicular to the F-N interface. 
(b) When the voltage is applied parallel to an ideal interface, no spin current is induced since $\nabla\cdot\Ev=0$.
\label{FIGspininjection}
}
\end{center}
\end{figure}

We have thus derived the explicit expression for the spin chemical potential within the linear response theory.
Let us apply \Eqref{muspin} to a ferromagnetic-normal metal junction with an insulating barrier, used in the nonlocal spin injection experiments  \cite{Kimura07}, depicted in Fig. \ref{FIGspininjection}(a).
When the voltage is applied perpendicular to the interface (we choose the $x$ axis in this direction), the electric field is uniform inside the ferromagnet and the normal metal except at the interface. 
Writing the voltage drop at the interface (chosen as at $x=0$) by $V_{\rm FN}$, we obtain 
\begin{align}
\nabla\cdot\Ev \simeq \delta(x) V_{\rm FN}/d,
\end{align}
 where $d$ is the width of the interaface, which is treated as small enough compared with the electron mean free path, resulting in the delta function in $\nabla\cdot\Ev$. 
In totally unpolarized non-magnetic metals, namely, if $\sigma_+=\sigma_-$ and $D_+=D_-$, the correlation function in \Eqref{chidef} always vanishes.
As is naively guessed, therefore, spin injection thus requires an effective spin polarization close to the interface, induced by the exchange interaction with the ferromagnet.
This spin polarization is expected to be localized within a short distance of a few lattice constants from the interface. 
Let us approximate the interface polarization by introducing spin-dependent diffusion constant and the density of states, $ \overline{D}_\sigma$ and $\overline\dos_\sigma$, respectively, at the interface.
The long-range behavior of the spin correlation function in the non-magnetic side is then obtained as
\begin{align}
\chi^{\rm (N)}(\xv) &=
-\frac{e}{4\pi} (\sum_\sigma \sigma \overline{D}_\sigma\overline\dos_\sigma)\frac{e^{-|\xv|/\ls{}}}{|\xv|}, 
\end{align}
where $\ls{}$ in the spin diffusion length in the normal metal 
( $\ls{}$ is a long ($\sim \mu$m) length scale and thus does not depend on the spin).
We therefore obtain from \Eqref{muspin} the chemical potential as
\begin{align}
\muspin^{\rm (N)} (\xv) &=
\frac{q_{\rm s} }{4\pi |\xv|}e^{-|\xv|/\ls{}},
\end{align}
 where 
$q_{\rm s}\equiv 
(\sum_\sigma \sigma \overline{D}_\sigma \overline\dos_\sigma) eV_{\rm FN})A_{\rm FN}/d
$,
 is the spin accumulation rate at the interface (per unit time), and $A_{\rm FN}$ is the area of the junction.
This result of $\muspin$ is consistent with intuitive and phenomenological results of the spin injection in the perpendicular structure shown in Fig. \ref{FIGspininjection}(a). 
In contrast, when the voltage is applied parallel to the ideal interface as shown in Fig. \ref{FIGspininjection}(b), spin injection does not occur since $\nabla\cdot\Ev=0$ at the interaface and thus $ \muspin^{\rm (N)}=0$.
%
%%%%%%%%%%%
%%%%%%%%%%%%%%

%\section{Discussion}

\section{Total torque and asymmetric $\beta$ term}

The continuity equation (\ref{sconteq}), indicates that the spin polarization (magnetization) changes due to the spin relaxation.
(Change of the magnetization magnitude is a feature of the itinerant magnetism.)
By use of Eqs. (\ref{TauE})(\ref{mudiffusioneq})(\ref{jsexpression2}), we see that \Eqref{sconteq}
results in
\begin{align}
\dot{s}^z 
&= \gamma_\torque  \nabla\cdot\Ev+\delta \gamma_\torque  \nabla_z E_z
+\sum_{\pm}(\pm)\frac{\mu_{\pm}}{(\ls{,\pm})^2}, \label{dotsz}
\end{align}
where 
$\gamma_\torque \equiv \gamma-\delta \sigma_{{\rm s}}$
and 
$\delta \gamma_\torque \equiv  \delta\gamma+\delta \sigma_{{\rm s}}$.
General case with uniform magnetization along any unit vector $\nv$ is given by ($\sv\equiv s\nv$) 
\begin{align}
\dot{\sv} 
&%= \nv(-\nabla\cdot \jspin{}{z}+{\cal T}^z)
=
\nv \lt( \gamma_\torque \nabla\cdot\Ev + \delta \gamma_\torque  \nabla_\parallel E_\parallel
+\sum_{\pm}(\pm)\frac{\mu_{\pm}}{(\ls{,\pm})^2}  \rt), \label{dotsnv}
\end{align}
where $E_\parallel\equiv \nv\cdot\Ev$ and $\nabla_\parallel\equiv \nv\cdot\nabla$.

In addition to the change of the magnitude, \Eqref{dotsnv}, there is a torque, which is perpendicular to $\nv$.
Such torque arises when the magnetization is not homogeneous, and plays important roles in current-induced magnetization dynamics. 
We have carried out the calculation of the current-induced torque done in Ref. \cite{TE08} on the same footing as the derivation of \Eqref{TauE}.
As a result, we found that the $\beta$ term becomes asymmetric as (see Appendix \ref{APP:torque} for details of the calculation)
\begin{align}
{\cal T}^{(\beta),\alpha}_{\so} 
& =
-\frac{P}{es^2} 
[\beta \sv\times (\jv\cdot\nabla)\sv 
+\delta \beta \sv\times (j_\parallel \nabla_\parallel)\sv]^\alpha,
\end{align} 
where $j_\parallel\equiv \nv\cdot\jv$ is the current along the local magnetization  
and $\delta \beta/\beta=-1/5$ in the present model.
The spin transfer torque due to the spin-orbit interaction is thus different from that due to the spin-flip scattering.
The expression of the total torque allowing for the spatially varying current density and the magnetization is therefore obtained as 
\begin{align}
\dot{\sv} 
&=  
-\frac{P}{2e}
(\nabla\cdot\jv)\nv 
-\frac{P}{e}\lt[ \beta \nv\times (\jv\cdot\nabla)\nv
+\delta \beta \nv\times (j_\parallel \nabla_\parallel)\nv)
\rt]
\nnr
& 
+ \nv
\lt(
\gamma_\torque (\nabla\cdot\Ev)
+\delta\gamma_\torque   (\partial_\parallel E_\parallel)
+\sum_{\pm} (\pm) (\ls{,\pm})^{-2}\mu_\pm
\rt).
\label{torquesum}
\end{align}
This expression clearly demonstrates that the spin relaxation torque requires some inhomogeneity either of the applied current or the spin structure,  in addition to the spin-orbit (or spin flip) interaction.
Totally homogeneous system does not relax.
The last term in Eq. (\ref{torquesum}) gives useful information for measuring the spin accumulation induced by the spin current.

\section{Conclusion}

We have carried out a microscopic calculation of the spin relaxation torque and the spin current induced in disordered ferromagnetic metals by the applied electric field. 
The spin-orbit interaction arising from the random impurities is included as a source of spin relaxation, and inhomogeneity of the applied electric field is taken into account.
We found that the spin relaxation torque in the uniform magnetization case is written as a divergence of the electric field plus an anisotropic term. 
The spin current was shown to be made up of field-driven (local) and diffusive (nonlocal) contributions, the latter written as a gradient of a spin chemical potential.
We have derived a general linear response expression for the spin chemical potential.
The spin injection effect was briefly discussed based on our results.
When the analysis is applied to the inhomogeneous magnetization case, we argued that the $\beta$ torque in the current-induced magnetization dynamics can be anisotropic.

Before finishing, we emphasize that the expression for the spin current and $\muspin$ are meaningless without specifying the physical observable to be measured.
In the inverse spin Hall effect, which was originally proposed as \cite{Hirsch99,Takahashi08,Saitoh06}
$j_\mu \propto \epsilon_{\mu\nu\rho}\jspin{,\nu}{\rho}$, 
it has recently been demonstrated that the charge current is not directly proportional to the spin current \cite{Hosono10,Takeuchi10}.
Solving for the spin current only does not therefore provide physical information.

\section*{Acknowledgment}
The authors thank E. Saitoh, J. Shibata, H. Kohno, S. Murakami for valuable discussions. 
This work was supported by a Grant-in-Aid for Scientific Research in Priority Areas, "Creation and control of spin current" (1948027), the Kurata Memorial Hitachi Science and Technology Foundation and the Sumitomo Foundation.

\appendix

\section{Derivation of the spin continuity equation}
\label{APP:eqofmo}

We calculate the commutators which appears in \Eqref{eqcommutators} with
the total Hamiltonian $H\equiv H_0+\Hso^0+\Hso^{A}+\Hem+\Himp$.
As for the free electron part, $H_0$, the commutator reads
\begin{align}
[H_0,c] &=   \frac{1}{2m}\nabla^2 c .  %\nonumber \\
%[H_0,c^\dagger] &=  -\frac{1}{2m}\nabla^2 c^\dagger.
\end{align}
The spin-orbit contributions read
\begin{align}
[\Hso^0,c] &=   i \sum_{ijk} \epsilon_{ijk} (\nabla_i \Vso^{(k)}) \sigma_k\nabla_j c    \nonumber \\
%[\Hso^0,c^\dagger]  &=   i \sum_{ijk} \epsilon_{ijk} (\nabla_i \Vso^{(k)}) (\nabla_j c^\dagger )\sigma_k,
[\Hso^{A},c] &=   e \sum_{ijk} \epsilon_{ijk} (\nabla_i \Vso^{(k)}) A_j \sigma_k c    ,
%[\delta \Hso,c^\dagger] &=  -e \sum_{ijk} \epsilon_{ijk} (\nabla_i \Vso^{(k)}) A_j  c^\dagger \sigma_k. 
\end{align}
and the electromagnetic field contribution is
\begin{align}
[\Hem,c] &=   -i\frac{e}{m} \sum_{i}  A_i \nabla_i c .
%[\Hem,c^\dagger] &=   -ie \sum_{i}  A_i \nabla_i c^\dagger,
\end{align}
The commutator for the creation operator is given by
$[H,c^\dagger]=-[H,c]^\dagger$. 
Equation (\ref{eqcommutators}) in the operator form  thus reads 
($\hat {s}^\alpha \equiv c^\dagger \sigma_\alpha c$ is the electron spin operator)
\begin{align}
\partial_t \hat{s}^\alpha
&=
\frac{i}{2m} \nabla \cdot (\cdag \sigma_\alpha \nablarl c) 
-\sum_{ij} \epsilon_{ij\alpha} (\nabla_i \Vso^{(\alpha)}) \nabla_j(\cdag c)
+\frac{e}{m} \sum_i A_i \nabla_i (\cdag \sigma_\alpha c)
\nnr
& 
+i\sum_{ijkl} \epsilon_{ijk} \epsilon_{ k\alpha l} (\nabla_i \Vso^{(k)})  (\cdag \sigma_l \nablarl_j c)
+2e \sum_{ijkl} \epsilon_{ijk} \epsilon_{k \alpha l} (\nabla_i \Vso^{(k)})  A_j (\cdag \sigma_l  c).
\label{eqofmoop}
\end{align}
(The contribution from the impurity scattering, $\Himp$,  vanishes.)
The first three terms on the right-hand side of \Eqref{eqofmoop} are written as divergence (choosing the electromagnetic vector potential as divergenceless, $\nabla\cdot\Av=0$).
From Eq. \ref{eqofmoop}, it is useful to define the spin current operators as
\begin{align}
\hat{j}_{\rm s}^{\alpha} \equiv \hat{j}_{\rm s}^{{\rm (n)},\alpha}+\hat{j}_{\rm s}^{{\rm so},\alpha}, 
\end{align}
where 
\begin{align}
 \hat{j}_{{\rm s},i}^{{\rm (n)},\alpha} 
  &\equiv -\frac{i}{2m} 
 \cdag \sigma^\alpha \vvec{\nabla}_i c 
 -\frac{e}{m}A_i  \cdag \sigma_\alpha c 
\nnr
\hat{j}_{{\rm s},i}^{\so, \alpha} 
 & \equiv 
- \sum_{j} \epsilon_{ij\alpha} 
(\nabla_j\Vso^{(\alpha)})  \cdag c.
\end{align}
Taking the average, the equation of motion for the spin density operator, \Eqref{eqofmoop},
 results in the spin continuity equation,
\begin{align}
\partial_t s^\alpha
&\equiv 
-\nabla \cdot ( \jsv^{{\rm (n)},\alpha}+\jsv^{{\rm so},\alpha})
+{\cal T}_{\so}^{\alpha} 
+{\cal T}_{\so}^{A,\alpha} ,
\end{align}
where each terms are given by Eqs. (\ref{spincurrentsdefs})(\ref{Tsodef}).

\section{Calculation of spin relaxation torque}
\label{SEC:appA}

In this section, we show details of calculation of spin relaxation torque.
The first contribution in Fig. \ref{FIGspinrelxationdiag1} reads 
\begin{align}
{\cal T}_{\so}^{z} 
& = 
i \frac{e}{m}\sumom\sumOm\sum_{\kv\kv'}
\sum_{\qv} \sum_{ijkl} \sum_{mn\beta\gamma}
\epsilon_{ijk} \epsilon_{z \beta k} 
\epsilon_{mn\gamma} e^{-i\qv\cdot\xv} e^{i\Omega t}
\average{ \Vso^{(k)}(\kv-\kv')  \Vso^{(\gamma)}(\kv'-\kv)} _{\rm i}
\nnr
&\times 
A_l 
(k-k')_i(k+k'+q)_j k'_m k_n \lt(k'+\frac{q}{2}\rt)_l \tr[\sigma_\beta g_{\kv,\omega} \sigma_\gamma 
g_{\kv',\omega} g_{\kv'+\qv,\omega+\Omega} ]^<
+{\rm c.c.}
\end{align}
Here $g_{\kv\omega}$ represents the contour ordered electron Green's function with the wave vector $\kv$ and the frequency $\omega$ ($g_{\kv\omega}$ is a $2\times2$ diagonal matrix in the spin space), tr is the trace over the spin, and  
$[\ ]^<$ represents the lesser components, and $\average{\ }_{\rm i}$ represents the averaging over the random spin-orbit impurities. 
Taking the lesser component, we obtain 
\begin{align}
{\cal T}_{\so}^{z} 
&=
-\frac{i}{2\pi}
\frac{e}{m}\sumOm\sum_{\kv\kv'}
\sum_{\qv} \sum_{ijkl} \sum_{mn\beta\gamma}
\epsilon_{ijk} \epsilon_{z \beta k} 
\epsilon_{mn\gamma} e^{-i\qv\cdot\xv} e^{i\Omega t}
\average{ \Vso^{(k)}(\kv-\kv')  \Vso^{(\gamma)}(\kv'-\kv)}_{\rm i} 
\nnr
&\times 
A_l \Omega 
(k-k')_i(k+k'+q)_j k'_m k_n \lt(k'+\frac{q}{2}\rt)_l \tr[\sigma_\beta \green{\kv}{\ret} \sigma_\gamma 
\green{\kv'}{\ret} \green{\kv'+\qv}{\adv} ]
+{\rm c.c.}
\end{align}
The retarded and advanced electron Green's functions at zero frequency are represented by $\green{}{\ret}$ and $\green{}{\adv}$, respectively.

The leading contribution for small $1/(\ef\tau)$ is given by
\begin{align}
{\cal T}_{\so}^{z} 
&=
-\frac{i}{\pi}
 \frac{e}{m}\sumOm\sum_{\kv\kv'}
\sum_{\qv} \sum_{ijkl} \sum_{mn\beta\gamma}
\epsilon_{ijk} \epsilon_{z \beta k} 
\epsilon_{mn\gamma} e^{-i\qv\cdot\xv} e^{i\Omega t}
\average{ \Vso^{(k)}(\kv-\kv')  \Vso^{(\gamma)}(\kv'-\kv)}_{\rm i} 
\nnr
&\times 
A_l \Omega 
k_i k'_j k'_m k_n k'_l \tr[\sigma_\beta \green{\kv}{\ret} \sigma_\gamma 
\green{\kv'}{\ret} \green{\kv'+\qv}{\adv} ]
+{\rm c.c.}
\end{align}
Expanding $\qv$ in the Green's function, we obtain the leading contribution as
\begin{align}
{\cal T}_{\so}^{z} 
&=
-\frac{i}{\pi}
 \frac{e}{m^2}\sumOm\sum_{\kv\kv'}
\sum_{\qv} \sum_{ijklo} \sum_{mn\beta\gamma}
\epsilon_{ijk} \epsilon_{z \beta k} 
\epsilon_{mn\gamma} e^{-i\qv\cdot\xv} e^{i\Omega t}
\average{ \Vso^{(k)}(\kv-\kv')  \Vso^{(\gamma)}(\kv'-\kv)}_{\rm i} 
\nnr
&\times 
A_l \Omega 
k_i k'_j k'_m k_n k'_l k'_o q_o \tr[\sigma_\beta \green{\kv}{\ret} \sigma_\gamma 
\green{\kv'}{\ret} (\green{\kv'}{\adv})^2 ]
+{\rm c.c.}
\end{align}
Using the rotational symmetry, i.e., 
$\average{k_i k_j}=\frac{k^2}{3}\delta_{ij}$ and 
$\average{k_i k_j k_l k_k}=\frac{k^4}{15}(\delta_{ij}\delta_{kl}+\delta_{ik}\delta_{jl}+\delta_{il}\delta_{jk})$ (the average here denotes the angular average), we obtain 
\begin{align}
{\cal T}_{\so}^{z} 
&=
-\frac{i}{\pi}
\frac{e}{m^2}\sumOm\sum_{\kv\kv'}
\sum_{\qv} \sum_{kl\beta\gamma}
\epsilon_{z \beta k} 
e^{-i\qv\cdot\xv} e^{i\Omega t}
\average{ \Vso^{(k)}(\kv-\kv')  \Vso^{(\gamma)}(\kv'-\kv)}_{\rm i} 
\nnr
&\times 
\frac{1}{45}
A_l \Omega k^2 (k')^4 (-4\delta_{k\gamma}q_{l}+\delta_{l\gamma}q_{k}+\delta_{kl}q_\gamma)
\tr[\sigma_\beta \green{\kv}{\ret} \sigma_\gamma 
\green{\kv'}{\ret} (\green{\kv'}{\adv})^2 ]
+{\rm c.c.}
\end{align}
 
We carry out the average over the impurity spin-orbit interaction as \Eqref{vsav} to obtain
\begin{align}
{\cal T}_{\so}^{z} 
&=
\frac{4i}{45\pi}
\nso \lamso^2 \frac{e}{m^2}\sumOm\sum_{\kv\kv'}
\sum_{\qv} \sum_{\beta\gamma}
\epsilon_{z \beta \gamma} 
e^{-i\qv\cdot\xv} e^{i\Omega t} \Omega  (\qv\cdot\Av -q_\gamma A_\gamma/2)
k^2 (k')^4 
\tr[\sigma_\beta \green{\kv}{\ret} \sigma_\gamma 
\green{\kv'}{\ret} (\green{\kv'}{\adv})^2 ]
+{\rm c.c.}
\end{align}

The asymmetric part of the trace is calculated by use of 
($A(\equiv \hf(A_++A_-+\sigma_z(A_+-A_-)))$ and $B(\equiv \hf(B_++B_-+\sigma_z(B_+-B_-)))$ are any diagonal $2\times2$  matrices)
\begin{align}
\tr[
(\sigma_\beta A \sigma_\gamma 
-\sigma_\gamma A \sigma_\beta )
B
]
&= -2i \epsilon_{\beta\gamma z} 
\sum_{\sigma}\sigma  A_\sigma B_{-\sigma} .
\end{align}
The result is
\begin{align}
{\cal T}_{\so}^{z} 
&=
\frac{2}{45\pi}
\nso \lamso^2 \frac{e}{m^2}\sumOm\sum_{\kv\kv'}
\sum_{\qv}
e^{-i\qv\cdot\xv} e^{i\Omega t} \Omega  
(3\qv\cdot\Av +q_z A_z)
k^2 (k')^4 
\sum_{\sigma} \sigma  \green{\kv,\sigma}{\ret} 
\green{\kv',-\sigma}{\ret} (\green{\kv',-\sigma}{\adv})^2 
+{\rm c.c.},
\end{align}
which is \Eqref{Tauzdominant}.

Since the summation over $\kv$ and $\kv'$ is dominated by the imaginary part of the Green's functions, we use
$\green{\kv,\sigma}{\ret} =-\frac{i}{2\tau_\sigma}|\green{\kv,\sigma}{\ret}|^2+o(\eftauinv)$ and 
$\green{\kv',-\sigma}{\ret} (\green{\kv',-\sigma}{\adv})^2 
=\frac{i}{2\tau_{-\sigma}}|\green{\kv',-\sigma}{\ret}|^4+o(\eftauinv)$, and obtain the leading contribution as
\begin{align}
{\cal T}_{\so}^{z} 
&=
- \frac{2\pi}{45}
\nso \lamso^2 \frac{e}{m^2}\sumOm 
\sum_{\qv}
e^{-i\qv\cdot\xv} e^{i\Omega t} \Omega  
(3\qv\cdot\Av +q_z A_z)
\dosu \dosd \kfu^2 \kfd^2 
(\kfu^2\tau_+^2-\kfd^2\tau_-^2).
\label{relaxationtorqueresultAPP}
\end{align}
Here spin-dependent density of states and the Fermi wavelength are represented by $\dos_\sigma$ and $\kfs$, respectively.

Another contribution to the relaxation torque, 
${\cal T}_{\so}^{A,\alpha} $ of \Eqref{TsoAdef}, represented by the third diagram of Fig. \ref{FIGspinrelxationdiag1},
reads 
\begin{align}
{\cal T}_{\so}^{A,\alpha}
& = 
i e\sumom\sumOm\sum_{\kv\pv}
\sum_{\qv\Omega} \sum_{ijk} \sum_{mn\beta\gamma}
\epsilon_{ijk} \epsilon_{\alpha \beta k} \epsilon_{mn\gamma} 
e^{-i\qv\cdot\xv} e^{i\Omega t}
\average{ \Vso^{(k)}(\pv)  \Vso^{(\gamma)}(-\pv)}_{\rm i} 
\nnr
&\times 
A_l (\qv,\Omega) 
(k-k')_i (k-k')_m (k+k')_n 
\tr[\sigma_\beta g_{\kv,\omega} \sigma_\gamma g_{\kv',\omega} ]^<_{\pv\equiv \kv'-\kv}.
\end{align}
As is seen, 
${\cal T}_{\so}^{A,\alpha}$ is an odd function of $k$ or $k'$, and thus it vanishes.

The second diagram in Fig .\ref{FIGspinrelxationdiag1} (we denote it as $\delta {\cal T}_{\so}^{\alpha}$)
 reads
\begin{align}
\delta {\cal T}_{\so}^{\alpha}
& = 
-e\sumom\sumOm\sum_{\kv\kv'}
\sum_{\qv\Omega} \sum_{ijk} \sum_{mn\beta\gamma}
\epsilon_{ijk} \epsilon_{\alpha \beta k} \epsilon_{mn\gamma} 
e^{-i\qv\cdot\xv} e^{i\Omega t}
\average{ \Vso^{(k)}(\kv'-\kv-\qv)  \Vso^{(\gamma)}(\kv-\kv'+\qv)}_{\rm i}  
\nnr
&\times 
A_l (\qv,\Omega) 
(k'-k-q)_i (k'-k-q)_n (k+k')_j 
\tr[\sigma_\beta g_{\kv,\omega} \sigma_\gamma g_{\kv',\omega+\Omega} ]^<.
\end{align}
Here, the wave vectors of the two Green's function, $\kv$ and $\kv'$, are independent,
 and thus the magnitude of $\delta {\cal T}_{\so}^{\alpha}$ turns out to be 
smaller than ${\cal T}_{\so}^{z}$ 
by the order of $1/(\ef \tau)$.

We therefore see that the dominant contribution to the relaxation torque arises from the first process in Fig. \ref{FIGspinrelxationdiag1}, which is estimated to be
\Eqref{relaxationtorqueresultAPP} (\Eqref{Tauzdominant}).

\section{Calculation of the spin current}
\label{SEC:appB}

\subsection{Normal spin current}

Here we show first the derivation of the normal spin current without the vertex correction, i.e., the first term of  \Eqref{spincurrent1}.
The leading contribution of \Eqref{spincurrentsdefs} arises from the first diagram of Fig. \ref{FIGspincurrent}. 
It reads
\begin{align}
\jspin{,i}{(0),z} 
&=
 i\frac{e}{m^2}
\sum_{\kv\qv}e^{-i\qv\cdot\xv}\sumOm \sumom e^{i\Omega t}
\sum_{j} k_i k_j 
A_j(\qv,\Omega) \sum_{\sigma} \sigma 
\nnr
&\times \lt[
\lt(f\lt(\ompOm\rt) -f\lt(\ommOm\rt)\rt)
\green{\kvmq,\ommOm,\sigma}{\ret} \green{\kvpq,\ompOm,\sigma}{\adv} 
\rt.\nnr
& \lt.
+f\lt(\ommOm\rt) 
\green{\kvmq,\ommOm,\sigma}{\adv} \green{\kvpq,\ompOm,\sigma}{\adv} 
-f\lt(\ompOm\rt) 
\green{\kvmq,\ommOm,\sigma}{\ret} \green{\kvpq,\ompOm,\sigma}{\ret} 
\rt]
\end{align}
Here the effect of spin-orbit interaction is included in the lifetime  (selfenergy) of the Green's functions. 
Estimating $\jspin{,i}{(0),z}$ to linear order in $\Omega$ and using $f'(\omega)=-\delta(\omega)$ at low temperatures, we see obviously that the first contribution in the square bracket reduces to the first term of  \Eqref{spincurrent1}.
The remaining contribution, which we call $\delta \jspin{,i}{(0),z}$, is calculated  to linear order in $\Omega$ as
\begin{align}
\delta\jspin{,i}{(0),z} 
&=
 i\frac{e}{m^2}
\sum_{\kv\qv}e^{-i\qv\cdot\xv}\sumOm \sumom e^{i\Omega t}
\sum_{j} k_i k_j 
A_j(\qv,\Omega) \sum_{\sigma} \sigma 
\nnr
&\times 
\lt[
f\lt(\omega\rt) \lt[ 
\green{\kvmq,\omega,\sigma}{\adv} \green{\kvpq,\omega,\sigma}{\adv} 
-\green{\kvmq,\omega,\sigma}{\ret} \green{\kvpq,\omega,\sigma}{\ret} 
\rt] \rt. \nnr
&
-\frac{\Omega}{2}f'\lt(\omega\rt) 
[
\green{\kvmq,\omega,\sigma}{\adv} \green{\kvpq,\omega,\sigma}{\adv} 
+\green{\kvmq,\omega,\sigma}{\ret} \green{\kvpq,\omega,\sigma}{\ret} 
] \nnr
&  
+\frac{\Omega}{2}f\lt(\omega\rt) 
[
(\green{\kvmq,\omega,\sigma}{\adv})^2 \green{\kvpq,\omega,\sigma}{\adv} 
-\green{\kvmq,\omega,\sigma}{\adv}(\green{\kvpq,\omega,\sigma}{\adv})^2 
\nnr 
& \lt.
-(\green{\kvmq,\omega,\sigma}{\ret})^2 \green{\kvpq,\omega,\sigma}{\ret} 
+\green{\kvmq,\omega,\sigma}{\ret}(\green{\kvpq,\omega,\sigma}{\ret})^2 
]
\rt]. \label{deltajs0}
\end{align}

On the other hand, the contribution from the gauge field, $\jspin{,i}{A,\alpha}$ in \Eqref{spincurrentsdefs}, reads
\begin{align}
\jspin{,i}{A,\alpha} (\xv,t)
&=
 i \delta_{\alpha,z} \frac{e}{m}A_i(\xv,t)
\sumom\sum_{\kv\sigma}
\sigma \green{\kv\omega\sigma}{<}
\nnr
&=
 i \delta_{\alpha,z} \frac{e}{m}A_i(\xv,t)
\sumom\sum_{\kv\sigma}
\sigma f(\omega)(\green{\kv\omega\sigma}{\adv}-\green{\kv\omega\sigma}{\adv})
\label{jsA1}
\end{align}
This contribution is proportional to $\Av$, and so is of the order of $\Ev/\Omega$.
One can easily see that this contribution cancels with the first term of \Eqref{deltajs0} estimated at $q=0$.
We therefore obtain 
\begin{align}
\delta\jspin{,i}{(0),z} +\jspin{,i}{A,\alpha}
&=
 i\frac{e}{m^2}
\sum_{\kv\qv}e^{-i\qv\cdot\xv}\sumOm \sumom e^{i\Omega t}
\sum_{j} k_i k_j 
A_j(\qv,\Omega) \sum_{\sigma} \sigma 
\nnr
&\times 
\lt[
-\frac{\Omega}{2}f'\lt(\omega\rt) 
[
\green{\kvmq,\omega,\sigma}{\adv} \green{\kvpq,\omega,\sigma}{\adv} 
+\green{\kvmq,\omega,\sigma}{\ret} \green{\kvpq,\omega,\sigma}{\ret} 
] \rt. \nnr
&  
+\frac{\Omega}{2}f\lt(\omega\rt) 
[
(\green{\kvmq,\omega,\sigma}{\adv})^2 \green{\kvpq,\omega,\sigma}{\adv} 
-\green{\kvmq,\omega,\sigma}{\adv}(\green{\kvpq,\omega,\sigma}{\adv})^2 
\nnr 
& \lt.
-(\green{\kvmq,\omega,\sigma}{\ret})^2 \green{\kvpq,\omega,\sigma}{\ret} 
+\green{\kvmq,\omega,\sigma}{\ret}(\green{\kvpq,\omega,\sigma}{\ret})^2 
]
\rt]+O(q). \label{deltajs1}
\end{align}
Since we consider a slowly varying gauge field, it is sufficient for our purpose to estimate this expression at $q=0$ in the Green's functions (the adiabatic limit).
The result is 
\begin{align}
\delta\jspin{,i}{(0),z} +\jspin{,i}{A,\alpha}
&=
 i\frac{e}{m^2}
\sum_{\kv\qv}e^{-i\qv\cdot\xv}\sumOm \sumom e^{i\Omega t}
\sum_{j} k_i k_j 
A_j(\qv,\Omega) \sum_{\sigma} \sigma 
\nnr
&\times 
\lt(-\frac{\Omega}{2}f'\lt(\omega\rt) \rt)
[
(\green{\kv,\omega,\sigma}{\adv})^2+(\green{\kv,\omega,\sigma}{\ret})^2 ] ,
\label{deltajs2}
\end{align}
which turns out after $\kv$-summation to be negligibly small 
(smaller by a factor of $1/(\ef \tau)$ than the dominant contribution).
We therefore obtain \Eqref{spincurrent1} as the dominant normal spin current at small $\Omega$ and $q$.

\subsection{Calculation of the third diagram in Fig. \ref{FIGspincurrent}}

The contribution from the third diagram in Fig. \ref{FIGspincurrent}, which arises from the normal current and the spin-orbit interaction including the gauge field, reads 
\begin{align}
\jspin{,i}{(3),\alpha} 
&=
-\frac{ie}{m} \sumom\sumOm\sum_{\kv\kv'}
\sum_{\qv}
\epsilon_{jk\beta} \epsilon_{mn\gamma} e^{-i\qv\cdot\xv} e^{i\Omega t}
\average{ \Vso^{(\beta)}(\pv)  \Vso^{(\gamma)}(-\pv)} _{\rm i} 
A_m(\qv,\Omega) 
 k_i(k-k')_n 
\nnr
&\times 
\lt[
\lt(k-\frac{q}{2}\rt)_j \lt(k'-\frac{q}{2}\rt)_k
\tr[\sigma^\alpha g_{\kvmq,\omega} \sigma_\beta  g_{\kvpmq,\omega} \sigma_\gamma g_{\kvpq,\omega+\Omega} ]
\rt.
\nnr
&+ \lt. 
\lt(k+\frac{q}{2}\rt)_j \lt(k'+\frac{q}{2}\rt)_k
\tr[\sigma^\alpha g_{\kvmq,\omega} \sigma_\gamma  g_{\kvppq,\omega+\Omega} \sigma_\beta g_{\kvpq,\omega+\Omega} ]   \rt]^<_{\pv=\kv'-\kv}.
\end{align}
Taking the lesser component, the dominant contribution at small $\Omega$ and $q$ turns out to be
\begin{align}
\jspin{,i}{(3),\alpha} 
&=
\frac{ie}{m} \sumom\sumOm\sum_{\kv\kv'}
\sum_{\qv}
\epsilon_{jk\beta} \epsilon_{mn\gamma} e^{-i\qv\cdot\xv} e^{i\Omega t}
\average{ \Vso^{(\beta)}(\kv'-\kv)  \Vso^{(\gamma)}(\kv-\kv')} _{\rm i} 
f'(\omega) \Omega A_m(\qv,\Omega) 
 k_ik_jk'_k k'_n 
\nnr
&\times 
\tr[\sigma^\alpha \gr_{\kv} (\sigma_\beta  \gr_{\kvp} \sigma_\gamma+\sigma_\gamma  \ga_{\kvp} \sigma_\beta)
 \ga_{\kv} ].
\end{align}
By use of the averaging over the spin-orbit potential diagonal in the spin space, \Eqref{vsav}, 
%$\average{ \Vso^{(\beta)}(\kv'-\kv)  \Vso^{(\gamma)}(\kv-\kv')} =\nso\lamso^2 \delta_{\beta\gamma}$, 
we obtain 
\begin{align}
\jspin{,i}{(3),\alpha} 
&=
-\frac{ie}{2\pi m} \nso\lamso^2 \sumOm\sum_{\kv\kv'}
\sum_{\qv}
\epsilon_{jk\beta} \epsilon_{mn\gamma} e^{-i\qv\cdot\xv} e^{i\Omega t}
\Omega A_m(\qv,\Omega) 
 k_ik_jk'_k k'_n 
\nnr
&\times 
\tr[\sigma^\alpha \gr_{\kv} \sigma_\beta  (\gr_{\kvp} + \ga_{\kvp}) \sigma_\gamma
 \ga_{\kv} ]_{\beta=\gamma}.
\end{align}
We therefore see by noting that 
$\sum_{\kvp} (\gr_{\kvp} + \ga_{\kvp})=O(\dos/(\ef\tau))$ that the contribution $\delta \jspin{,i}{\so,\alpha} $
is negligibly small compared with the main contribution, \Eqref{spincurrent1}.

\subsection{Anomalous spin current}

The anomalous spin current arising from the spin-orbit interaction, defined in \Eqref{jssodef} and shown in Fig. \ref{FIGspincurrent} (the fourth and fifth diagrams), 
is calculated in a similar manner as that for the relaxation torque. 
The dominant contribution arises from the fourth diagram of Fig. \ref{FIGspincurrent}.
It reads
\begin{align}
\jspin{,i}{\so,\alpha} 
&\simeq \jspin{,i}{(4),\alpha}
\nnr
& \equiv
-\frac{e}{m}\sumom\sumOm\sum_{\kv\kv'}
\sum_{\qv}
\epsilon_{ij\alpha} \epsilon_{mn\gamma} e^{-i\qv\cdot\xv} e^{i\Omega t}
\average{ \Vso^{(\alpha)}(\kv-\kv')  \Vso^{(\gamma)}(\kv'-\kv)}_{\rm i}  
\nnr
&\times 
A_l 
(k-k')_j k'_m k_n \lt(k'+\frac{q}{2}\rt)_l 
\tr[g_{\kv,\omega} \sigma_\gamma  g_{\kv',\omega} g_{\kv'+\qv,\omega+\Omega} ]^<    .
\end{align}
Taking the lesser component and the average over the spin-orbit potential, and by noting that only $\gamma=z$ contribution is finite because of the trace over spin, we obtain
\begin{align}
\jspin{,i}{\so,z} 
&=
\frac{1}{2\pi}
\nso\lamso^2 \frac{e}{m}\sumOm\sum_{\kv\kv'}
\sum_{\qv}
\epsilon_{ijz} 
\epsilon_{mnz} e^{-i\qv\cdot\xv} e^{i\Omega t}
A_l \Omega 
(k-k')_j k'_m k_n \lt(k'+\frac{q}{2}\rt)_l \tr[ \green{\kv}{\ret} \sigma_z 
\green{\kv'}{\ret} \green{\kv'+\qv}{\adv} ]
\end{align}
The leading contribution at small $q$ is then obtained as
\begin{align}
\jspin{,i}{\so,z} 
&=
\frac{-i}{2\pi}
\nso\lamso^2 \frac{e}{m} \dot{A}_i(\xv,t)  
\sum_{\kv\kv'}
\epsilon_{ijz} 
(k_j)^2 (k'_i)^2 \tr[ \green{\kv}{\ret} \sigma_z 
\green{\kv'}{\ret} \green{\kv'}{\adv} ]
\nnr
&=
\lt\{
\begin{array}{cc}
-\frac{\pi}{9}
\nso\lamso^2 \frac{e}{m} \dot{A}_i(\xv,t)  
\sum_{\sigma}\sigma (\kfs)^4 (\dos_{\sigma})^2 \tau_\sigma 
& (i=x,y) \\
0 & (i=z)\end{array}
\rt., \label{jssores}
\end{align}
where we used the rotational symmetry in $\kv$ space.
This expression is \Eqref{jssores2}.

The last diagram of Fig. \ref{FIGspincurrent} reads
\begin{align}
\jspin{,i}{(5),\alpha} 
=
\delta_{\alpha z} \frac{ie}{2\pi}
\nso\lamso^2 
\epsilon_{ijz}  
\epsilon_{klz} 
\dot{A}_l \sum_{\kv\kv'}
(k-k')_j (k-k')_k 
\tr[ \green{\kv}{\ret} \sigma_z 
\green{\kv'}{\adv} ].
\end{align}
This contribution turns out to be smaller than \Eqref{jssores} by the order of $(\ef\tau)^{-1}$.

\section{Calculation of asymmetric $\beta$ torque}
\label{APP:torque}
The torque generated by the spin-orbit interaction and uniform electric field when the spin structure is inhomogeneous was calculated in Ref. \cite{TE08}.
There, the angular averaging over the electron wave vectors was carried out assuming rotational symmetry on averaging the spin-orbit interaction, resulting in a symmetric torque called the $\beta$ term. 
Here we carry out the calculation without such approximation, and show that asymmetric $\beta$ term then arises.
The torque reads (see Eqs. (19)(21)(35) of Ref. \cite{TE08})
\begin{align}
{\cal T}^\alpha_{\so} 
&=
-\frac{e}{\pi m^2} \nso\lamso^2 \sum_{\mu\nu\beta\gamma ij}\epsilon_{\alpha\mu\nu}R_{\mu\beta}R_{\nu\gamma}
E_i A_j^\delta \sum_{\kv\kv'} k_i k_j [(\kv\times\kv')_\nu]^2  
\nnr
& \times 
\tr[(\sigma_\beta \gr_{\kv'}\sigma_\gamma+\sigma_\gamma \ga_{\kv'}\sigma_\beta)
(|\gr_{\kv}|^2 \sigma_\delta \ga_{\kv}+\gr_{\kv} \sigma_\delta |\gr_{\kv}|^2) ],  \label{tauasymmetric}
\end{align}
where $R_{\mu\nu}\equiv 2m_\mu m_\nu-\delta_{\mu\nu}$ denotes a rotation matrix for the gauge transformation and 
$ A_j^\delta\equiv (\mv\times\nabla_j\mv)^\delta$ is the spin gauge field 
($\mv\equiv (\sin\frac{\theta}{2}\cos\phi,\sin\frac{\theta}{2}\sin\phi,\cos\frac{\theta}{2})$, where 
$\theta$ and $\phi$ are polar coordinates representing the magnetization direction) \cite{TE08}.
In Ref. \cite{TE08}, the term $[(\kv\times\kv')_\nu]^2$ arising from the spin-orbit interaction was averaged independent of the other factor, $k_ik_j$, and thus the torque was proportional to  
$\average{k_ik_j}=\frac{k^2}{3}\delta_{ij}$ (Eq. (37) of Ref. \cite{TE08}). 
Here we carry out the averaging without assuming special symmetry. 
The averaging over the wave vectors in \Eqref{tauasymmetric} then reads   
\begin{align}
\average { k_i k_j [(\kv\times\kv')_\nu]^2 }%|_{\nu \mbox{\rm : fixed}}
&=
\frac{2}{45}k^4(k')^2(2\delta_{ij}-\delta_{i\nu}\delta_{j\nu}).
\end{align}
The last term gives rise to asymmetric torques.
Evaluating the trace in the spin space, the torque reads
\begin{align}
{\cal T}^\alpha_{\so} 
&=
\frac{8e}{45 \pi m^2} \nso\lamso^2 \sum_{\mu\nu\beta\gamma}\epsilon_{\alpha\mu\nu}R_{\mu\beta}R_{\nu\gamma}
\lt(\Ev\cdot\Av^\delta-\hf E_\nu A_\nu^\delta\rt) 
\sum_{\kv\kv'} k^4 (k')^2 
\nnr
& \times \sum_{\sigma}
\lt[
(\epsilon_{\beta\gamma\delta}-\epsilon_{\beta\gamma z}\delta_{\delta z}) 
|\gr_{\kv\sigma}|^2 \Re \gr_{\kv,-\sigma}\sum_{\sigma'}\Im \gr_{\kv'\sigma'} 
\rt. \nnr
& 
+2\epsilon_{\beta\gamma z}\delta_{\delta z}
|\gr_{\kv\sigma}|^2 \Re (\gr_{\kv\sigma}) \Im \gr_{\kv',-\sigma}
\nnr
& \lt.
+ (\delta_{\beta z}\delta_{\gamma \delta}-\delta_{\gamma z}\delta_{\beta \delta})
\sigma |\gr_{\kv\sigma}|^2 \Im \gr_{\kv,-\sigma}\sum_{\sigma'}\Im \gr_{\kv'\sigma'} 
\rt]. \label{tauasymmetric2}
\end{align}
Symmetric contribution (contribution from the $\Ev\cdot\Av^\delta$ term)
to the first line in the square bracket 
was calculated in Ref. \cite{TE08} by use of 
\begin{align}
\sum_{\mu\nu\beta\gamma}
\epsilon_{\alpha\mu\nu}R_{\mu\beta}R_{\nu\gamma} A_i^\delta 
(\epsilon_{\beta\gamma\delta}-\epsilon_{\beta\gamma z}\delta_{\delta z}) 
&=
2(A_i^\alpha +n^\alpha A_i^z) 
= (\nv\times \nabla_i\nv)^\alpha,
\end{align}
where $\nv$ denotes the magnetization direction. 
(
The second and third terms in the square bracket in \Eqref{tauasymmetric2} lead to a correction to the spin Berry phase 
(proportional to $\Av^z$) and the standard spin transfer torque 
 (proportional to $\nabla \nv^\alpha$), respectively.
)
As for the asymmetric contribution (terms proportional to $\hf E_\nu A_\nu^\delta$ in \Eqref{tauasymmetric2}), summation over $\nu$ cannot taken independent of the electric field, resulting in complicated torques.
We here look into only the correction to the $\beta$ term (the first line in the square bracket), which is obtained as (neglecting irrelevant terms)
\begin{align}
-\hf \sum_{\mu\nu\beta\gamma}
\epsilon_{\alpha\mu\nu}R_{\mu\beta}R_{\nu\gamma} E_\nu A_\nu^\delta 
(\epsilon_{\beta\gamma\delta}-\epsilon_{\beta\gamma z}\delta_{\delta z}) 
& \simeq 
\hf(\Ev\cdot\Av^\alpha-E_\alpha A_\alpha^\alpha) .
\end{align}
The result of the $\beta$ term is thus 
\begin{align}
{\cal T}^{(\beta),\alpha}_{\so} 
& \simeq 
\frac{P}{e} 
[\beta \nv\times (\jv\cdot\nabla)\nv 
+\delta \beta \nv\times (j_\parallel \nabla_\parallel)\nv],
\end{align} 
where $j_\parallel\equiv \nv\cdot\jv$,  
$\nabla_\parallel\equiv \nv\cdot\nabla$,
\begin{align}
\beta \equiv 
\frac{8e}{9 \pi m^2 \sigma_B} \nso\lamso^2 \sum_{\kv\kv'}\sum_{\sigma\sigma'}
k^4(k')^2 |\gr_{\kv\sigma}|^2 \Re \gr_{\kv,-\sigma} \Im \gr_{\kv'\sigma'} ,
\end{align}
($\sigma_B$ is the Boltzmann conductivity) 
and $\delta \beta/\beta=-1/5$ in the present model.
Thus the $\beta$ term arising from the spin-orbit interaction has asymmetric contribution, in contrast to the one arising from the random magnetic impurities \cite{KTS06}.

%\bibliography{/home/tatara/References/09,/home/tatara/References/gt}

\end{document}